\documentclass[iop]{emulateapj}

\newcommand{\twelveco}{\mbox{$^{12}$CO }} 
\newcommand{\thirteenco}{\mbox{$^{13}$CO }} 
\newcommand{\twelvecoh}{\mbox{$^{12}$CO $J$=2--1 }} 
 
\newcommand{\twelvecol}{\mbox{$^{12}$CO $J$=1--0 }} 
\newcommand{\thirteencol}{\mbox{$^{13}$CO $J$=1--0 }} 
\newcommand {\msun}{\mbox{$M_\odot$ }}
\newcommand {\msune}{\mbox{$M_\odot$}}
\newcommand {\msunpyr}{\mbox{$M_\odot$ yr$^{-1}$ }}
\newcommand {\kms}{\mbox{km~s$^{-1}$ }}
\newcommand {\kmse}{\mbox{km~s$^{-1}$}}

\newcommand {\rtwoone}{\mbox{$R_{2-1/1-0}$ }}

\newcommand{\twelvecohe}{\mbox{$^{12}$CO $J$=2--1}}

\newcommand{\thirteencole}{\mbox{$^{13}$CO $J$=1--0}}

\slugcomment{}

\shorttitle{Discovery of possible molecular counterparts of the DHN}
\shortauthors{R. Enokiya et al.}

\begin{document}


\title{Discovery of possible molecular counterparts to the infrared Double Helix Nebula in the Galactic center}


\author{R. Enokiya\altaffilmark{1}, K. Torii\altaffilmark{1}, M.Schultheis\altaffilmark{2}, Y.Asahina\altaffilmark{3}, R. Matsumoto\altaffilmark{3}, E.Furuhashi\altaffilmark{1}, K.Nakamura\altaffilmark{1}, K.Dobashi\altaffilmark{4}, S.Yoshiike\altaffilmark{1}, J.Sato\altaffilmark{1}, N.Furukawa\altaffilmark{1}, N.Moribe\altaffilmark{1}, A.Ohama\altaffilmark{1}, H.Sano\altaffilmark{1}, R.Okamoto\altaffilmark{1}, Y.Mori\altaffilmark{1}, N.Hanaoka\altaffilmark{1}, A,Nishimura\altaffilmark{5}, T.Hayakawa\altaffilmark{1}, T.Okuda\altaffilmark{1}, H.Yamamoto\altaffilmark{1}, A.Kawamura\altaffilmark{1,6}, N.Mizuno\altaffilmark{1,6}, T.Onishi\altaffilmark{1,5}, M.R.Morris\altaffilmark{7}, and Y. Fukui\altaffilmark{1}}
\affil{$^1$Department of Physics, Nagoya University, Furo-cho, Chikusa-ku, Nagoya, Aichi, 464-8602, Japan}
\affil{$^2$ Universit\'e de Nice Sophia-Antipolis, CNRS, Observatoire de C\^ote d'Azur, Laboratoire Lagrange, BP4229, 06304 Nice Cedex 4, France}
\affil{$^3$Faculty of Science, Chiba University, Inage-ku, Chiba 263-8522}
\affil{$^6$National Astronomical Observatory of Japan, Mitaka, Tokyo, 181-8588, Japan}
\affil{$^5$Department of Astrophysics, Graduate School of Science, Osaka Prefecture University, 1-1Gakuen-cho, Nakaku, Sakai, Osaka 599-8531, Japan}
\affil{$^4$Department of Astronomy and Earth Sciences, Tokyo Gakugei University, Koganei 594-1, Tokyo 184-8501, Japan}
\affil{$^7$Department of Physics and Astronomy, University of California, Los Angeles, California 90095-1547, USA.}

\email{enokiya@a.phys.nagoya-u.ac.jp}




\begin{abstract}
We have discovered two molecular features at radial velocities of  --35 \kms and 0 \kms toward the infrared Double Helix Nebula (DHN) in the Galactic center with NANTEN2. The two features show good spatial correspondence with the DHN. We have also found two elongated molecular ridges at these two velocities distributed vertically to the Galactic plane over 0\fdg8. The two ridges are linked by broad features in velocity and are likely connected physically with each other. The ratio between the \twelvecoh and $J$=1--0 transitions is 0.8 in the ridges which is larger than the average value 0.5 in the foreground gas, suggesting the two ridges are in the Galactic center. An examination of the $K$ band extinction reveals a good coincidence with the CO 0 \kms ridge and is consistent with a distance of 8$\pm$2 kpc. We discuss the possibility that the DHN was created by a magnetic phenomenon incorporating torsional Alfv\' en waves launched from the circum--nuclear disk \citep{mor2006} and present a first estimate of the mass and energy involved in the DHN.

\end{abstract}



\section{Introduction}
The Galactic center (GC) harbors a massive black hole, Sgr A*, and is still a most enigmatic region in the Galaxy after a few decades of observational studies at various wavelengths as well as theoretical studies. The highly complex distribution of various astronomical objects and the lack of optical images due to the heavy extinction make it difficult to disentangle the observed features into individual physical entities and to establish their interrelations (e.g., Morris $\&$ Serabyn 1996).

In spite of these difficulties it is becoming increasingly evident that the magnetic field plays an important role in the GC region. \citet{yus1984} discovered the Radio Arc, in which synchrotron emission arises in highly ordered magnetic field lines vertical to the Galactic plane. Subsequently, more than 20 smaller non-thermal filaments were discovered to extend vertically to the Galactic plane within the central 300 pc of the GC \citep{mor1985,lar2004,yus2004}. Recently, based on molecular observations, \citet{fuk2006} discovered two huge molecular loops within $5^\circ$ of the GC and advocated that they are magnetic structures created by the Parker instability \citep{par1966}. Follow-up studies have shown details of the physical properties of the molecular loops, including shock heating of the footpoints of the loops \citep{mac2009,fuj2009,tak2009,tor2010a,tor2010b,kud2011}. At infrared wavelengths, $Spitzer$ images were used to identify the Double Helix Nebula (DHN), located $0\fdg6$ to $0\fdg8$ above Sgr A* \citep{mor2006}. The DHN appears to have two intertwined helical strands reminiscent of deoxyribonucleic acid. \citet{mor2006} proposed that the DHN is driven by torsional Alfv\'en waves that may be connected to the circumnuclear disk (CND) around Sgr A*. On the other hand, another possibility was discussed that the DHN may be associated with the eastern part of the GC lobe \citep{sof1984}, which may be connected to the Radio Arc but not to Sgr A*, based on polarization measurements of 10 GHz and 5 GHz radio continuum radiation \citep{tsu2010,law2008}. Most recently, the break in the non--thermal radio spectrum in the GC was used to infer a typical large--scale magnetic field of at least 50 $\mu$G within 200 pc of the GC \citep{cro2010}. These observational studies indicate that the magnetic field may play a crucial role in the gas kinematics in the GC.

Among these features, we shall here focus on the DHN. The previous study of the DHN is only at infrared wavelengths and the mass and energy involved in the DHN have not been estimated from observations. The basic physical properties of the DHN  therefore remains elusive. In order to better understand its physical properties and possible connection with the relevant features, we have carried out a search for molecular counterpart(s) in the rotational transitions of CO toward the Central Molecular Zone (CMZ, Morris $\&$ Serabyn 1996) with the 4m NANTEN2 telescope. The observations cover a large area of the CMZ with angular resolutions of 90$''$ and 180$''$, enabling us to make a direct comparison with the infrared distributions of the DHN. In this study we present discovery of possible molecular counterparts to the DHN and discuss their implications for the activity in the GC.  A fuller account of these observations will be published separately (R. Enokiya et al. 2013b, in preparation). The present paper is organized as follows. Section 2 describes the CO observations and Section 3 gives the results and an analysis of the near--infrared extinction. Section 4 gives comparisons with other wavelengths and an overall discussion and Section 5 summarizes the paper.

\section{Observations}
Observations of the $J$=2--1 transition of $^{12}$CO at 230 GHz were made during the period from 2010 July to 2011 January with the NANTEN2 4-m submillimeter/millimeter telescope located at 4865 m above the sea level in Atacama, Chile. The total observing time was $\sim$250 hr. The half power beamwidth (HPBW) was 90$''$ as measured by observing Jupiter. The telescope was equipped with a 4 K cooled superconducting--insulator--superconducting (SIS) mixer receiver with a double sideband (DSB) system temperature, including the atmosphere toward the zenith, of $\sim$180 K. 
Spectroscopy was done with a digital spectrometer whose bandwidth and resolution were 1 GHz and 61 kHz, respectively. The velocity coverage and resolution were 1300 km s$^{-1}$ 
and 0.08 km s$^{-1}$, respectively. Observations were made in the on-the-fly (OTF) mode and the final rms noise fluctuations are 0.35 K for a velocity resolution of 0.65 km s$^{-1}$ after convolving the spectral data with a Gaussian function.

Deep observations of the $J$=1--0 transition of $^{13}$CO at 110 GHz toward $|l| <$ 0\fdg25 and -0\fdg5 $<$ b $<$ 1\fdg0 were carried out with NANTEN2 by using the OTF mode during 2011 April to June. A 4 K cooled SIS mixer receiver provided a typical system temperature of $\sim$270 K in DSB, and a digital spectrometer provided a bandwidth and resolution of 1 GHz and 61 kHz, which correspond to 2600 km s$^{-1}$ and 0.17 km s$^{-1}$, respectively. We smoothed the obtained data to a velocity resolution of 0.51 \kms and angular resolution of 190$''$, and finally obtained rms noise fluctuations of $\sim$0.13 K for that velocity resolution.

Large--scale observations of the $J$=1--0 transition of $^{12}$CO and $^{13}$CO emission toward $|l| <$ 10\fdg0 and $|b| <$ 1\fdg0 were carried out with NANTEN2 by using the OTF mode during a period from 2011 May to 2013 January. Typical system temperature including the atmosphere toward the zenith was 200 K to 270 K in DSB, and a digital spectrometer provided a bandwidth and resolution of 1 GHz and 61 kHz. We smoothed the obtained data to an angular resolution of 200$''$, and finally obtained rms noise fluctuations of 0.5 K -- 1.0 K per channel for $^{12}$CO and 0.30 K -- 0.35 K per channel for $^{13}$CO.
 The velocities in the present paper always refer to the local standard of rest (LSR).

\section{Results}
\subsection{Discovery of candidate molecular counterparts to the DHN}
Figure \ref{ii_all} shows the integrated intensity distributions of \twelvecoh emission in an area of 4 degrees in l by 2 degrees in b centered at (l, b) = (0\fdg0, 0\fdg0) obtained with NANTEN2. The main components of the CMZ are distributed at $|b| <$ 0\fdg5. The position of the DHN is shown by a box toward (l, b) = (0\fdg0, 0\fdg75) in this figure. We show the velocity channel distributions of the \twelvecoh emission from the full region in Figures \ref{ap_channel1}--\ref{ap_channel3} of the Appendix. By analyzing the velocity structure in these figures, we have found two possible molecular counterparts to the DHN at two different velocities.

Figures \ref{lb}a and \ref{lb}b show the candidate counterparts of the DHN at velocity near -35 \kms and 0 km~s$^{-1}$, respectively (hereafter, the -35 \kms and 0 \kms features, respectively). These two linear features are both aligned with the DHN, and extend up to b = 0\fdg86 (the -35 \kms feature) and b = 0\fdg83 (the 0 \kms feature). The -35 \kms feature abuts the DHN on its western side, while the 0 \kms feature is centered along the eastern edge of the DHN. Both have widths of 10 pc -- 20 pc, if they are located at the distance of the GC, 8 kpc. We also made observations toward these two CO features with higher spatial resolutions of $\sim33''$, by using the CSO 10m and Mopra 22m telescopes in the CO $J$=2--1 and $J$=1--0 transitions, and found that the 0 \kms feature apparently coincides with the DHN and that the -35 \kms feature is spatially anti--correlated with the DHN. These results reinforce the association of the two CO features with the DHN and will be presented separately (K. Torii et al. 2013a, submitted). 

Figures \ref{meanv}a and \ref{meanv}b show the detailed velocity distributions of the two features and indicate that they have similar velocity gradients of $\sim 0\fdg4$ \kms pc$^{-1}$ in the east--west direction in the same sense. We estimate the molecular masses of the -35 \kms (0\fdg6 $\le$ b $\le$ 0\fdg86) and 0 \kms features (0\fdg6 $\le$ b $\le$ 0\fdg83) to be $4\times 10^3$ \msun  and $1\times 10^4$ \msune, respectively, for an assumed X factor of $0.7\times 10^{20}$ cm$^{-2}$/(K~\kmse) \citep{tor2010b} at 8 kpc.

\subsection{Molecular ridges rising from the Galactic plane}
We have found two other CO components apparently connected to the CO--emitting candidate DHN counterparts and we describe their distribution in this subsection.

\begin{figure*}
.pdfscale{1.05}
\includegraphics[scale= 1.05]{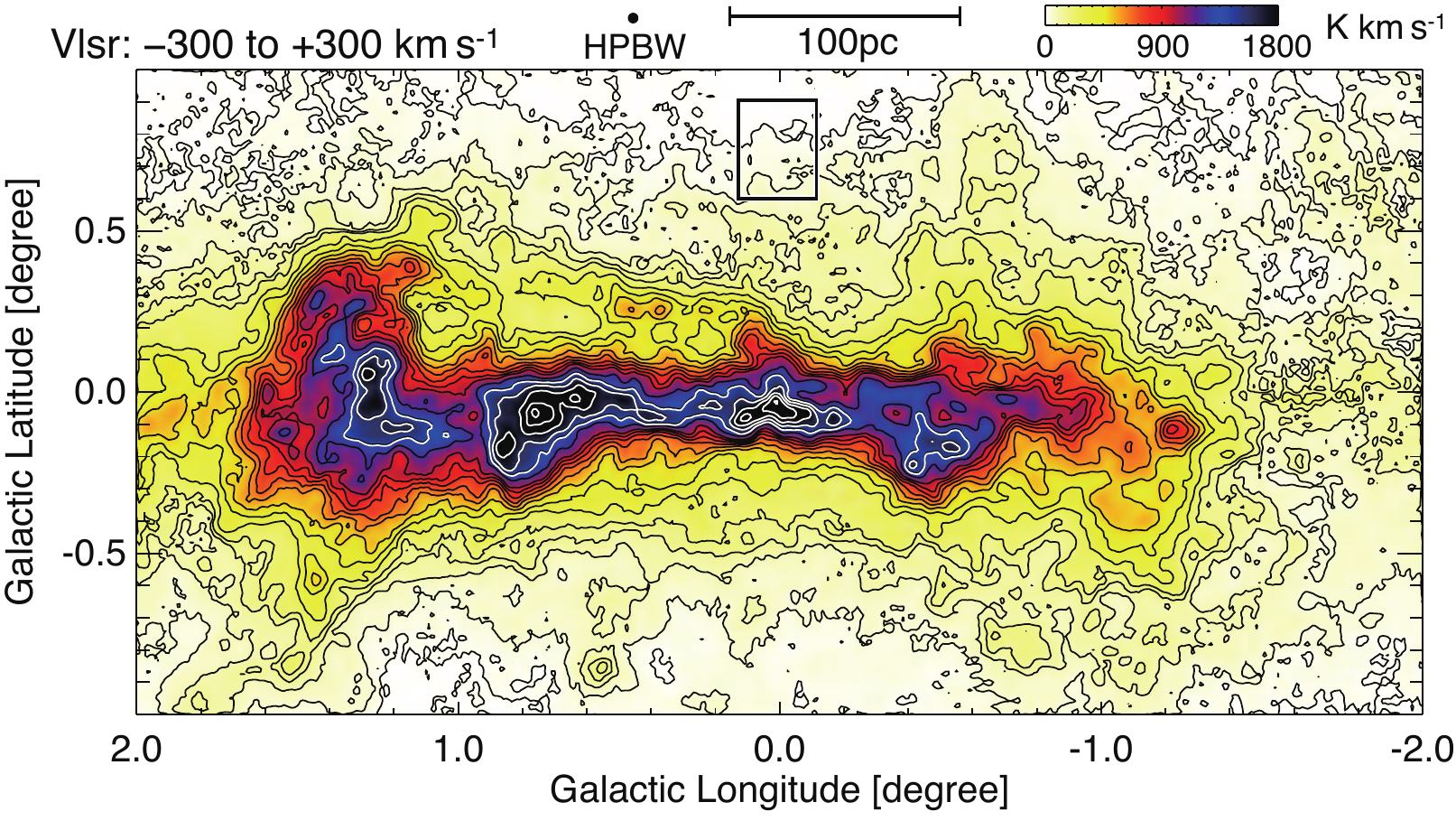}
\caption{$^{12}$CO $J$=2--1 distribution integrated over the velocity range of $-300$ to +300 km s$^{-1}$. Contours are drawn at 15, 50, 100, 150, 200, 300, 400, 500, 600, 700, 800, 900, 1000, 1200, 1400, 1600, 1800 and 2000 K km s$^{-1}$. A box shows the displayed area of Figure \ref{lb}.\label{ii_all}}
\end{figure*}

\begin{figure*}
.pdfscale{1.0}
\includegraphics[scale= 1.00]{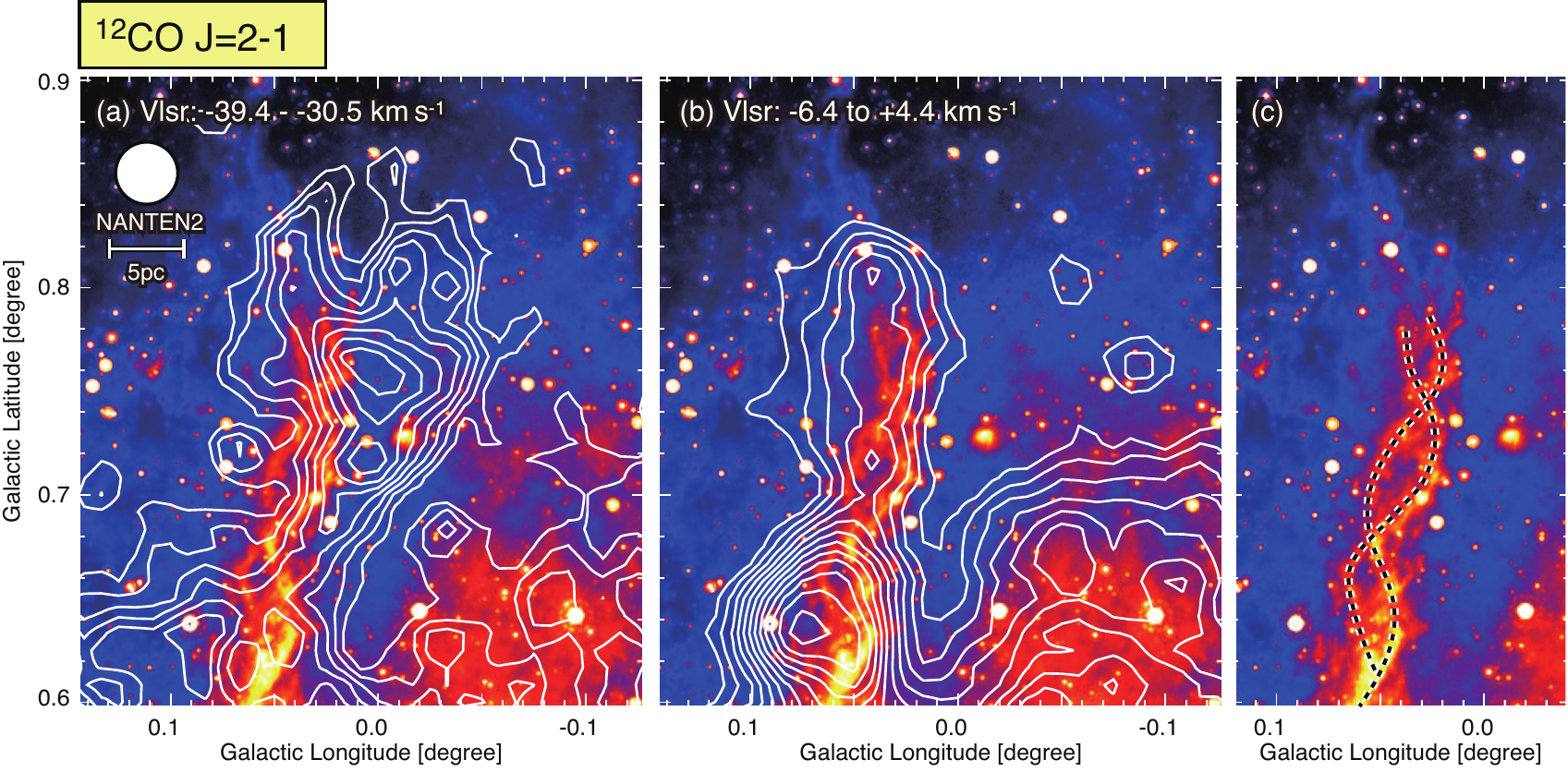}
\caption{(a--b) $^{12}$CO $J$=2--1 integrated intensity distributions of the two CO counterparts of the DHN superposed on the $Spitzer$ 24 $\mu$m image. (a) $-35$ km s$^{-1}$ feature. Contours are plotted at every 0.7 K km s$^{-1}$ from 1.0 K km s$^{-1}$. (b) 0 km s$^{-1}$ feature. Contours are plotted at every 1.4 K km s$^{-1}$ from 10 K km s$^{-1}$. (c) The $Spitzer$ 24 $\mu$m image with schematic lines of the two helical patterns.\label{lb}}
\end{figure*}

\begin{figure*}
.pdfscale{1.0}
\includegraphics[scale= 1.20]{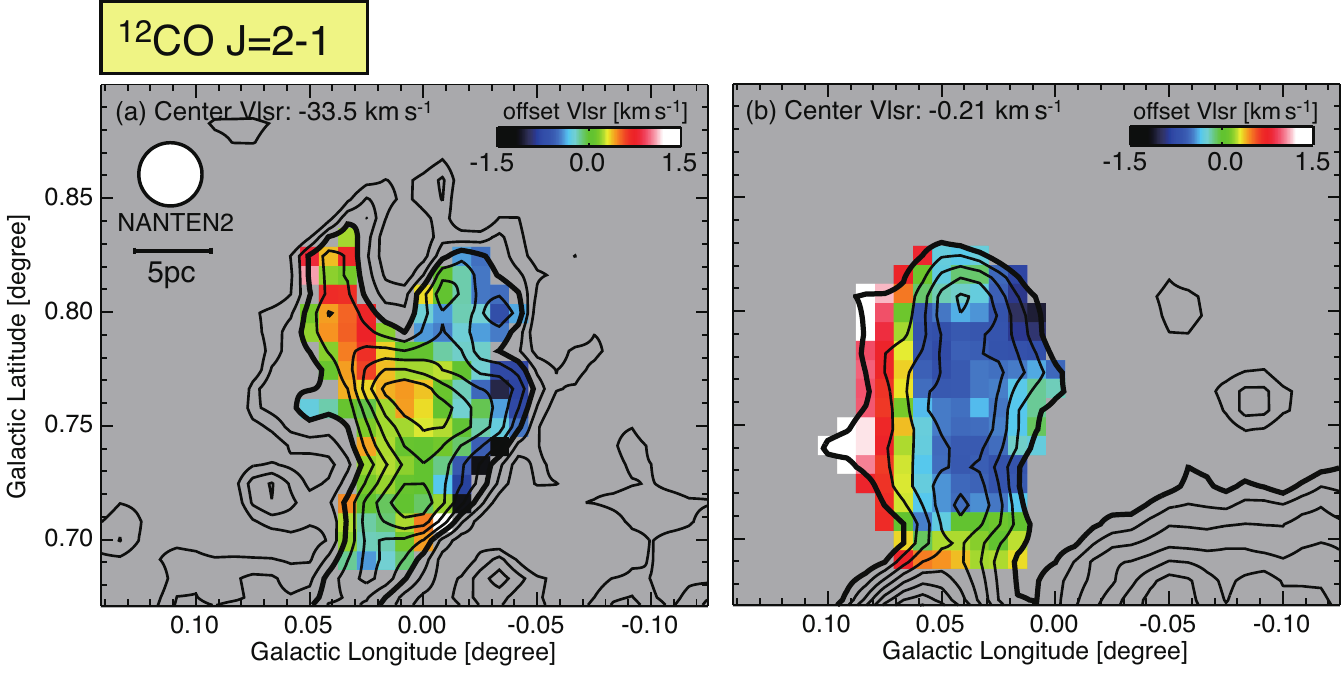}
\caption{Distributions of the intensity-weighted averaged velocity of the $-35$ km s$^{-1}$ and 0 km s$^{-1}$ features. The images show velocity offsets from their central velocities, $-33.5$ km s$^{-1}$ and $-0.21$ km s$^{-1}$. Contours shows the integrated $^{12}$CO $J$=2--1 emission and are plotted at the same levels as in Figures \ref{lb}a and \ref{lb}b.\label{meanv}}
\end{figure*}

\begin{figure*}
.pdfscale{.95}
\includegraphics[scale= .95]{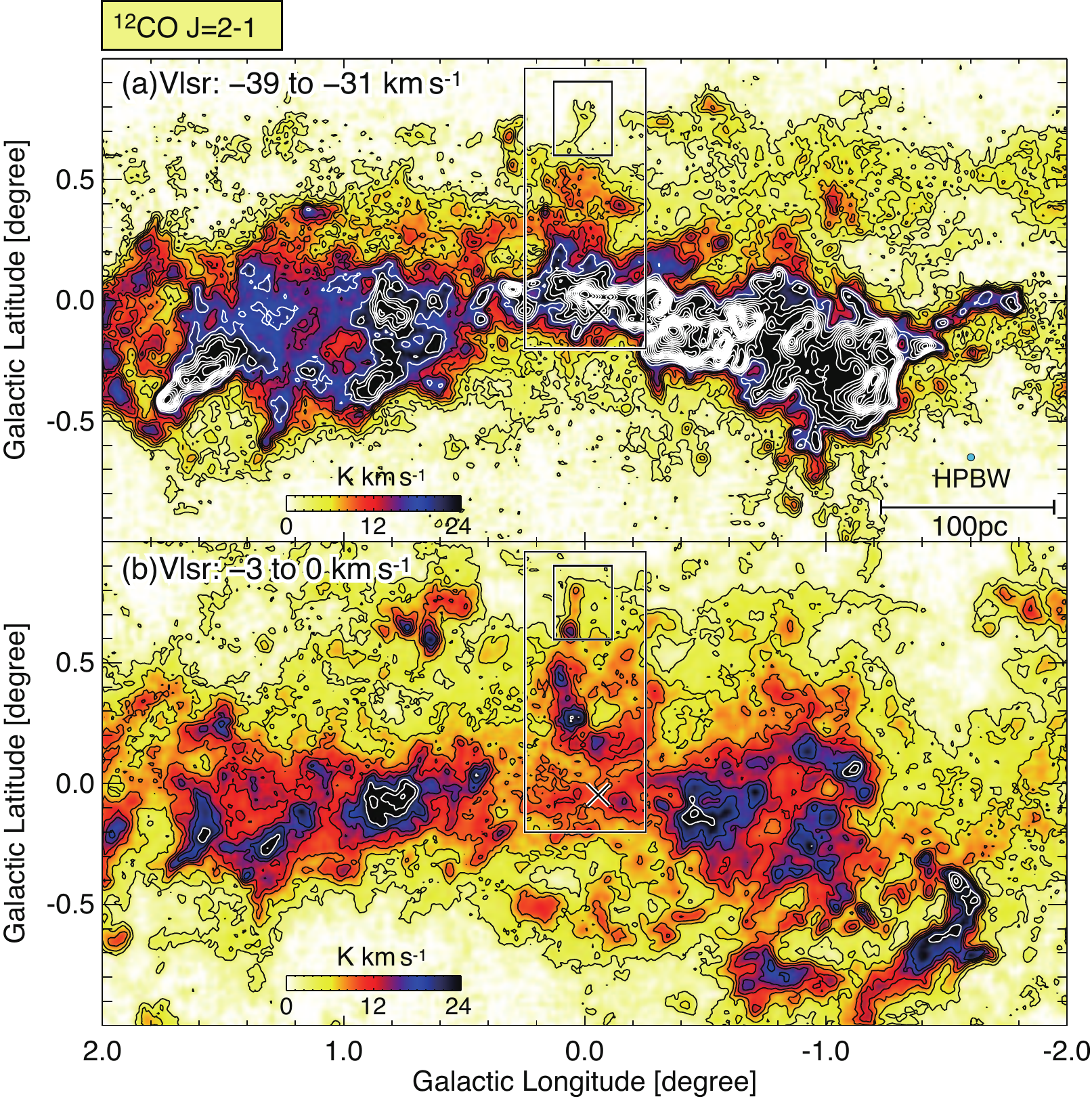}
\caption{Integrated intensity distributions of $^{12}$CO $J$=2--1 emission toward (l, b) = (-2\fdg0 to +2\fdg0, -1\fdg0 to +1\fdg0). A cross indicates the position of SgrA*, and big boxes and small boxes show the displayed regions of Figures \ref{lb} and \ref{lb13}, respectively. (a) Map of the $-35$ km s$^{-1}$ feature integrated over the velocity range from -31 to -39 \kmse. Contours are drawn every 3 K km s$^{-1}$ from 3 K km s$^{-1}$ (black lines) and every 6 K km s$^{-1}$ from 24 K km s$^{-1}$ (white lines). (b) Map of the 0 \kms feature integrated over the velocity range from -3 to 0 \kmse. Contours are drawn every 3 K km s$^{-1}$ from 3 K km s$^{-1}$ and every 6 K km s$^{-1}$ from 24 K km s$^{-1}$.\label{lb_large}}
\end{figure*}

\begin{figure*}
.pdfscale{1.05}
\includegraphics[scale= .95]{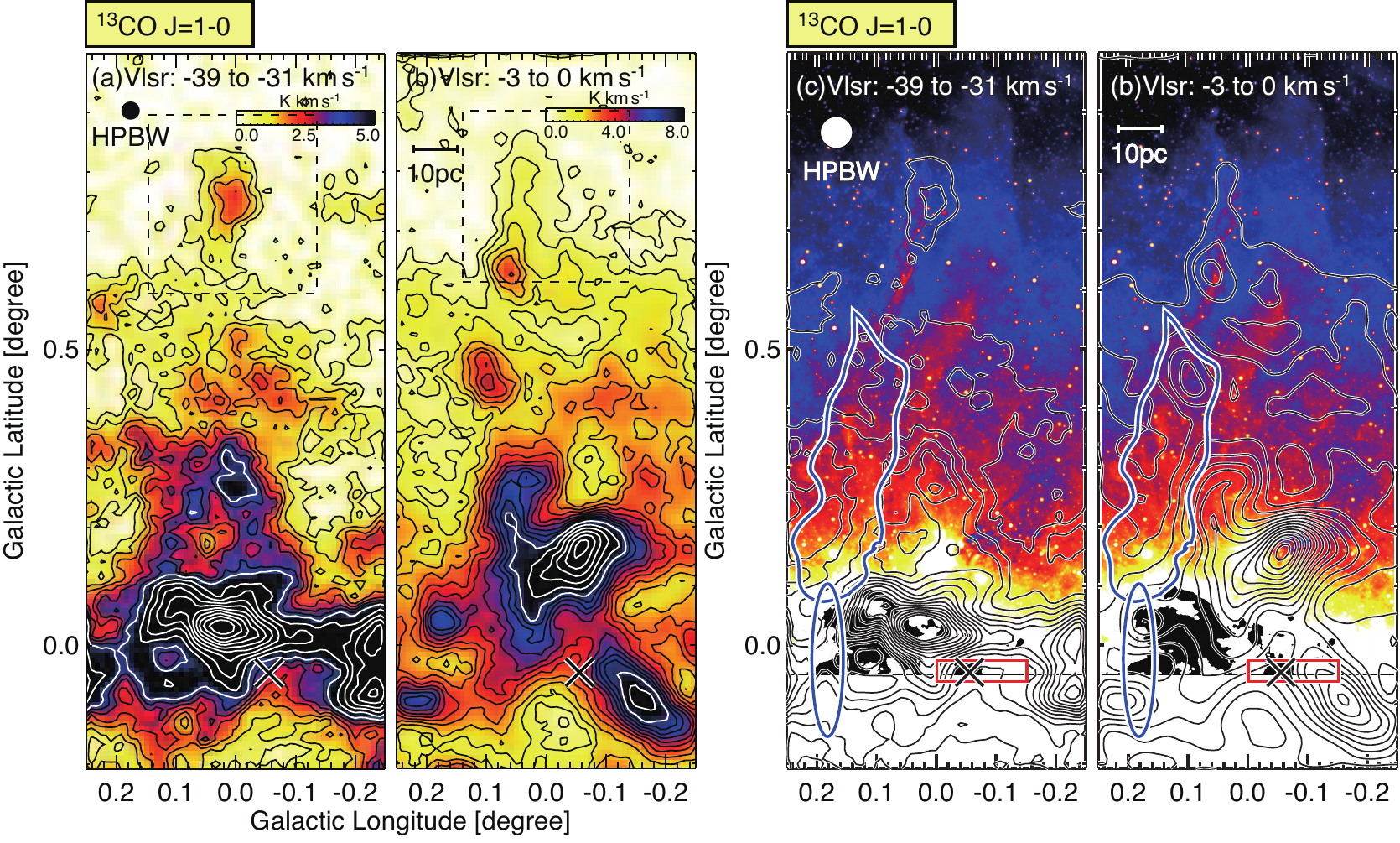}
\caption{(a)--(b) Integrated intensity distributions of the $^{13}$CO $J$=1--0 emission toward (l, b) = (-0\fdg25 to +0\fdg25 , -0\fdg2 to +1\fdg0). Crosses depict the location of SgrA*. (a) Map of the $-35$ km s$^{-1}$ feature integrated over the velocity range from -31 to -39 \kmse. Contours are drawn every 0.6 K km s$^{-1}$ from 0.6 K km s$^{-1}$ and then every 1.5 K km s$^{-1}$ from 4.2 K km s$^{-1}$. (b) Map of the $0$ km s$^{-1}$ feature integrated over the velocity range from -3 to 0 \kmse. Contours are drawn every 0.6 K km s$^{-1}$ from 0.6 K km s$^{-1}$ and then every 1.2 K km s$^{-1}$ from 7.2 K km s$^{-1}$. (c)--(d) Superpositions of smoothed contours of \thirteencol and the $Spitzer$ 24 $\mu$m image. Red box and blue ellipse give the approximate extent of the Radio Arc and CND, respectively. Blue contour is 50 mJy beam$^{-1}$ of linearly polarized intensity at 10 GHz, showing the approximate extent of the polarized plume. (c) $-35$ km s$^{-1}$ feature and ridge. Contours are plotted at every 1.5 K km s$^{-1}$ from 1.5 K km s$^{-1}$. (d) 0 km s$^{-1}$ feature and ridge. Contours are plotted at every 0.8 K km s$^{-1}$ from 1.2 K km s$^{-1}$. \label{lb13}}
\end{figure*}

Figures \ref{lb_large} shows the large--scale \twelvecoh distribution, including the two DHN counterpart features located in the small box. In Figure \ref{lb_large}a, we find that there is a ridge--like CO feature in the large box (hereafter, the -35 \kms ridge) just below the -35 \kms feature, extending toward the Galactic plane (at l = -0\fdg1 -- 0\fdg2). The -35 \kms ridge cannot be distinguished at latitudes $<$ 0\fdg2 because of confusion with the extended emission along the Galactic plane. Most of the $^{12}$CO emission in Figure \ref{lb_large}a is from the CMZ, since the distribution is similar to that of the main body of the CMZ (Figure \ref{ii_all}). 
In Figure \ref{lb_large}b the 0 \kms feature is clearly seen in the small box. In Figure \ref{lb_large}b we find a narrow ridge component (at l = 0\fdg0 -- 0\fdg2) at latitudes below $\sim$0\fdg6 (hereafter, the 0 \kms ridge), vertically extending down to b $\sim$ 0\fdg2 (in the large box). It is located just below the 0 \kms feature and has a width a few times larger than that of the 0 \kms feature. This ridge is affected by self--absorption by foreground gas below b = 0\fdg2, while the optically thin \thirteencol emission (Figure \ref{lb13}b) shows that the 0 \kms ridge extends to b = 0\fdg0.  Some of the $^{12}$CO features evident at higher Galactic latitudes in Figure \ref{lb_large}b, e.g., those at (l, b) = (-1\fdg7 to -0\fdg9, -1\fdg0 to -0\fdg2), (-1\fdg0 to -0\fdg6, -1\fdg0 to -0\fdg5) and (0\fdg4 to 0\fdg9, 0\fdg5 to 0\fdg8), are foreground features outside the GC, while the main CO emission in the region (l, b) = (-1\fdg2 to +1\fdg8, -0\fdg5 to +0\fdg4) is mostly from the CMZ. The $^{12}$CO feature at (l, b) = (0.3, 0.5 to 0.7) having velocity around -35 \kms may possibly be associated with the -35 \kms ridge, although no significant enhancement in the $^{12}$CO 2--1/1--0 intensity ratio of the feature (Figure 10a) does not strongly support the association.

Figures \ref{lb13}a--\ref{lb13}b show the \thirteencol emission toward the two ridges. In Figure \ref{lb13}a the -35 \kms ridge has a peak in the Galactic plane at l = 0\fdg0, whereas its distribution below b = 0\fdg1 is not clear due to contamination with the CMZ. Above b = 0\fdg15,  the -35 \kms ridge has two features in l = -0\fdg1 to 0\fdg0 and in l = 0\fdg05 to 0\fdg15 at b = 0\fdg2 that apparently merge at (l, b) = (0\fdg0, 0\fdg3). Beyond 0\fdg35 we see continuation of $^{13}$CO up to 0\fdg55. The 0 \kms ridge is seen in Figure \ref{lb13}b. The strong peak toward (l, b) = (-0\fdg05, 0\fdg20) is foreground, as shown by the near infrared extinction (see Figures \ref{ext}a and \ref{ext}d). Except for this feature we see the ridge at b = -0\fdg1 to 0\fdg3 and an additional peak at b = 0\fdg45. Figures \ref{lb13}c and \ref{lb13}d show a comparison with the $Spitzer$ infrared image. We note that the CO ridges show spatial correspondence with the infrared distribution in l = -2\fdg0 to 2\fdg0 and b above 0\fdg3, where contamination is not so heavy. In particular, the correspondence is good for the infrared DHN and its filament directed toward the Galactic plane (Figure \ref{lb13}). 
We estimate the molecular masses of the -35 \kms and the 0 \kms ridges to be $2\times10^5$ \msun and $1\times10^5$ \msune, respectively, for an assumed X factor $0.7\times 10^{20}$ cm$^{-2}$/(K~\kmse) \citep{tor2010b} at 8 kpc. Detailed spatial distributions of these ridges are shown in Figures \ref{ap_channel4} and \ref{ap_channel5} of the Appendix.

\subsection{Velocity distribution of the CO ridges}
\begin{figure*}
\begin{center}
.pdfscale{.8}
\includegraphics[scale= 1.1]{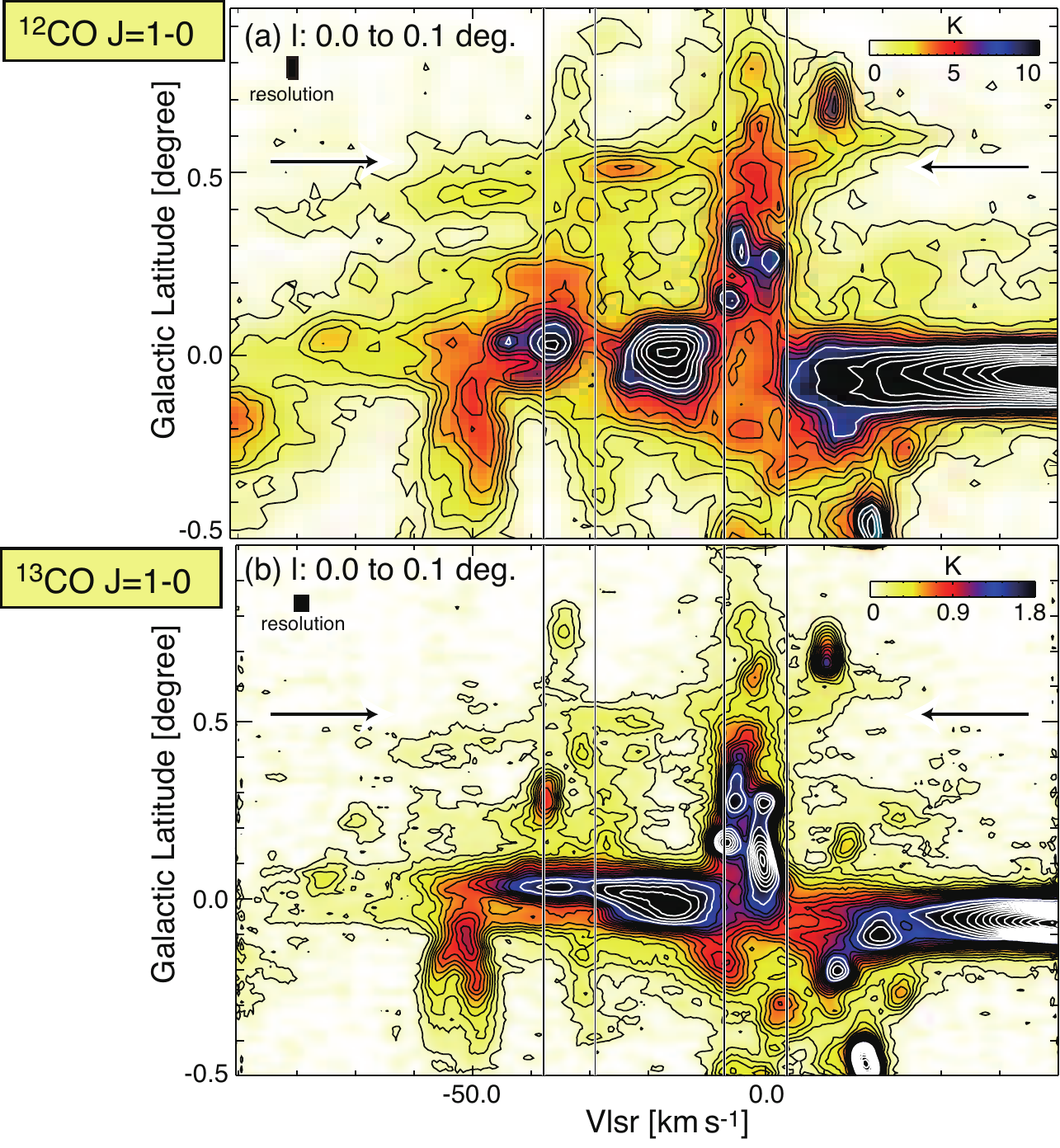}
\caption{Position-velocity distributions of averaged CO intensities toward the $-$35 km s$^{-1}$ and 0 km s$^{-1}$ ridges. The integration range in galactic longitude is $0\fdg0$ to $+0\fdg1$. (a) $^{12}$CO $J$=1--0. Contours are plotted at every 0.7 K from 0.5 K (black) and every 2.0 K from 7.5 K (white). Vertical lines delineate the two ridges, and arrows indicate the broad feature connecting the two ridges. (b) $^{13}$CO $J$=1--0. 
Contours are plotted at every 0.10 K from 0.08 K (black) and then every 0.2 K from 1.48 K (white).\label{pv}}
\end{center}
\end{figure*}

\begin{figure*}
.pdfscale{1.}
\includegraphics[scale= 0.90]{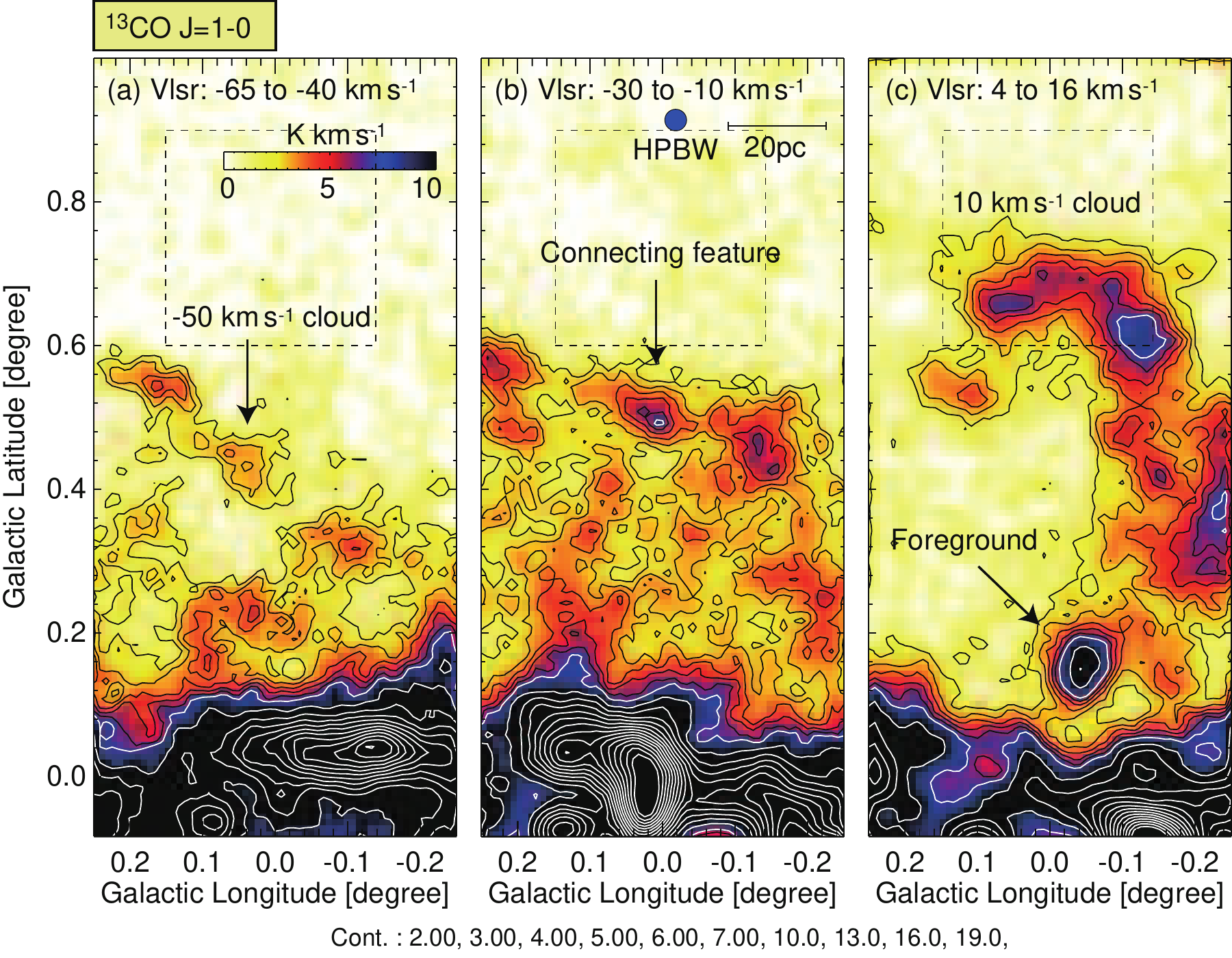}
\caption{Integrated intensity maps of \thirteencol emission in different velocity ranges. Contour values and integrated velocity ranges for each map are shown at the bottom and top of each figure, respectively. Dashed boxes show the same region of Figure \ref{lb}, where the DHN is distributed.\label{cnct}}
\end{figure*}

Figures \ref{pv}a and \ref{pv}b show the velocity--latitude distributions of the \twelvecol and \thirteencole, respectively, where the integration range in galactic longitude is from 0\fdg0 to 0\fdg1. The two CO ridges appear as narrow features with a FWHM line width of 5 -- 10 \kmse. The -35 \kms ridge appears at b = 0\fdg2 to 0\fdg35 with a velocity span of  $\la$ 20 \kms and the 0 \kms ridge at b = 0\fdg2 to 0\fdg6 with a velocity span of 15 \kmse. In Figure \ref{pv}a we find significant self--absorption features at velocities of 0 \kmse, -30 \kms and -50 \kms at $|b| \la$ 0\fdg2, while they are not seen in the optically thin $^{13}$CO in Figure \ref{pv}b. The self--absorption is attributable to the foreground low--excitation gas. The 0 \kms CO ridge extends to the Galactic plane at b = 0\fdg0. The strongest broad feature seen at b less than 0\fdg2 is the CMZ. There is another broad feature which appears to be connecting the -35 \kms and 0 \kms ridges at b = 0\fdg5, as indicated by black arrows in Figure \ref{pv} (see also Figures \ref{cnct}b and \ref{schepv}b). There is also weak emission from 35 \kms to 0 \kms filling the region below b = 0\fdg5 between the two CO ridges.

We present more details of the connecting feature between the two ridges in Figure \ref{cnct}, which shows the integrated intensity distributions of the \thirteencol emission in three different velocity ranges; (a) the -50 \kms cloud in a velocity range of -65 \kms to -40 \kmse, (b) the connecting feature in a velocity range of -30 \kms to -10 \kmse, and (c) the 10 \kms cloud in a velocity range of 4 \kms to 16 \kmse. In Figure \ref{cnct}b, the peak located at (l, b) = (0\fdg0, 0\fdg5) is that of the connecting feature and other two peaks at (l, b) = (-0\fdg12, 0\fdg45) and (0\fdg25, 0\fdg55) are also possibly related to the connecting feature (see Figure \ref{ap_channel2} in the Appendix). The -50 \kms cloud is extended from (l, b) = (-0\fdg1, 0\fdg3) to (0\fdg14, 0\fdg55) with a tilt of 40 degrees to the Galactic plane and the 10 \kms cloud shows a arced shape elongated in l at b = 0\fdg7 with extension toward the Galactic plane at l = -0\fdg1. The connecting feature seems to be unrelated to the -50 \kms (a) and 10 \kms clouds (c), since the three show significantly different spatial distributions. 

\begin{figure*}
\begin{center}
.pdfscale{.8}
\includegraphics[scale= .80]{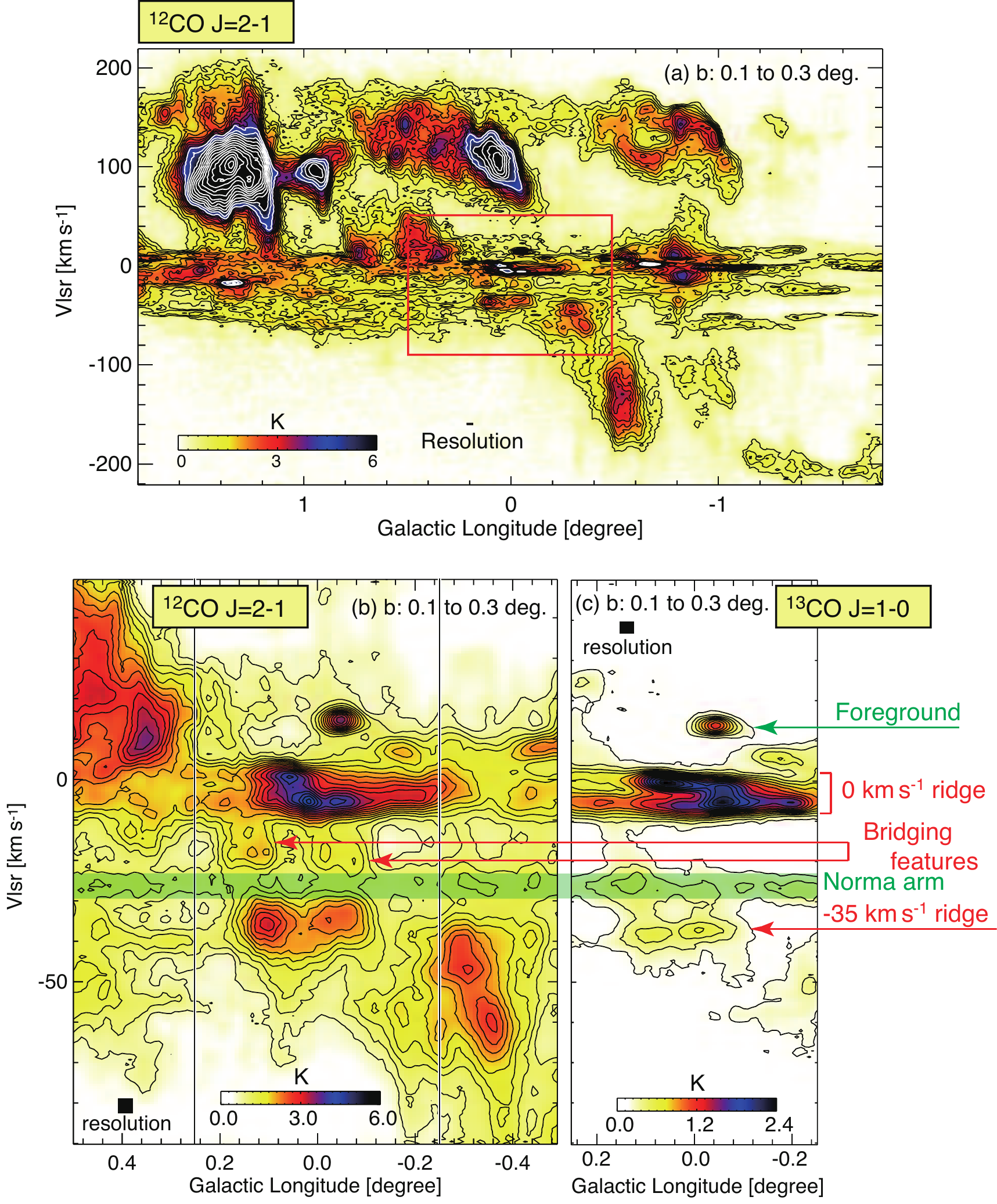}
\caption{Position--velocity distributions of averaged CO intensities toward the $-$35 km s$^{-1}$ and 0 km s$^{-1}$ ridges in the latitude range $0\fdg1 \le b \le 0\fdg3$. (a) Large scale longitude -- velocity diagram of \twelvecohe. (b) Close up view of the red box in panel a. Contours are plotted at every 0.25 K from 0.75 K. Vertical lines show the region shown in panel \ref{pv2}c. (c)$^{13}$CO $J$=1--0. Contours are plotted at every 0.15 K from 0.2 K.  \label{pv2}}
\end{center}
\end{figure*}

\begin{figure*}
\begin{center}
.pdfscale{0.8}
\includegraphics[scale= .80]{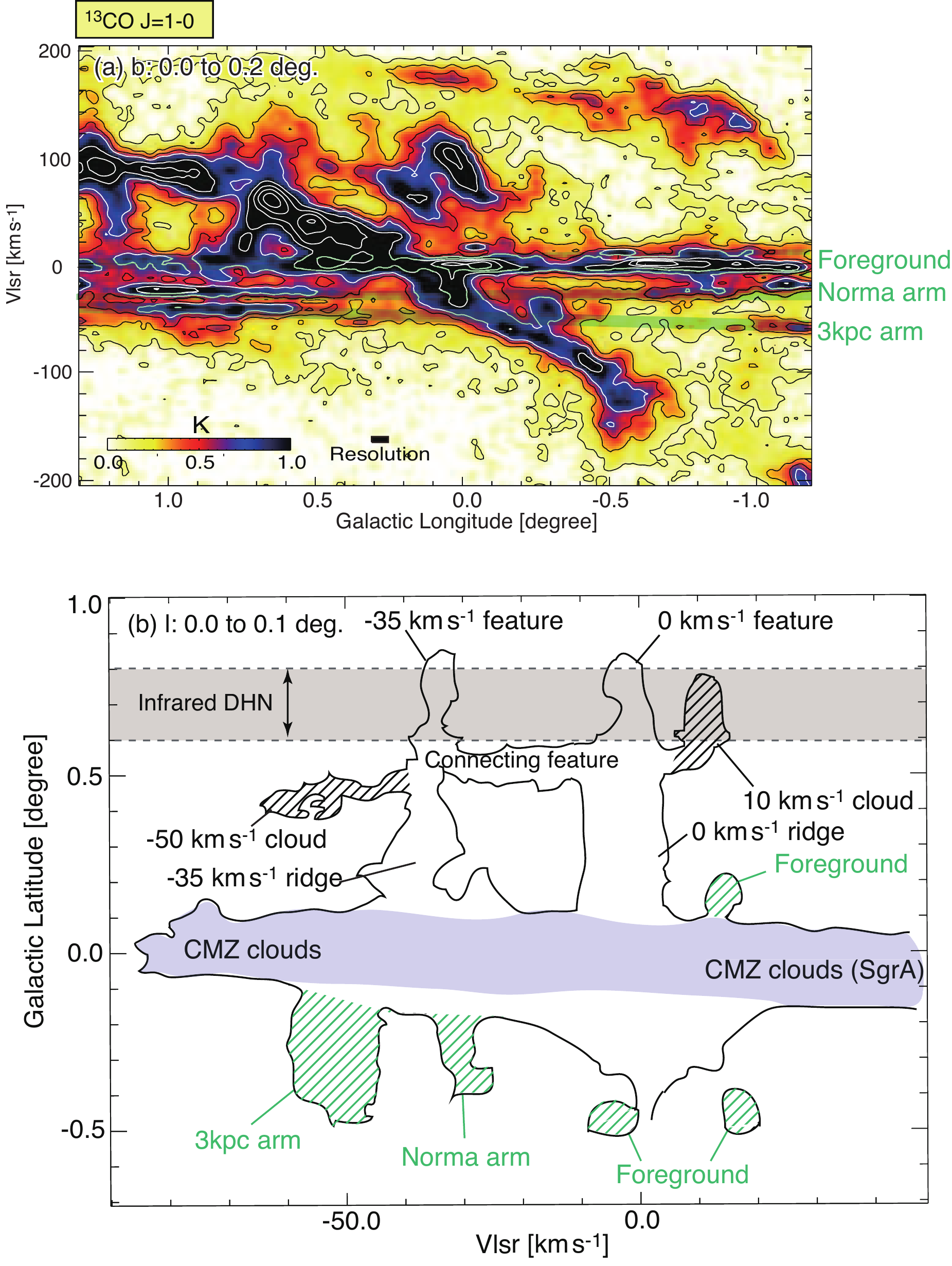}
\caption{(a) Longitude-velocity diagrams of $^{13}$CO $J$=1--0 integrated from -0\fdg03 to +0\fdg20. Green lines show emission from foreground clouds. (b)Schematic image of Figure \ref{pv}b. Green hatching and gray belt mean foreground cloud and dense molecular gas located in the CMZ, respectively.\label{schepv}}
\end{center}
\end{figure*}

The CO distribution toward the two CO ridges is complicated due to heavy contamination and self--absorption by different features in the line of sight, especially near V{lsr} = 0 \kmse. It is not straightforward to disentangle the contamination by using only the two--dimensional distribution. Figure \ref{pv2}a shows a large scale longitude--velocity diagram of \twelvecoh emission integrated between b = 0\fdg1 to 0\fdg3. The continuous CO feature which is distributed from (l, v) = (-0\fdg6, -150 \kmse) to (0\fdg2, 30 \kmse) is ascribed to part of the CMZ, because the velocity range is similar to that of the CMZ \citep{sof1995,tsu1999,mol2011}. Figures \ref{pv2}b and \ref{pv2}c show close--up views of the red box in Figure \ref{pv2}a in \twelvecoh and \thirteencole. Figure \ref{pv2}b shows that the -35 \kms CO ridge continues as two bridging features at l = -0\fdg05 and 0\fdg1 to the CO ridge at 0 \kmse, while it shows an intensity depression at -30 \kms due to the line absorption by the Norma arm. This distribution supports the hypothesis that the two ridges are connected by the two bridging features, and that the l = 0\fdg1 bridging feature, which has a $^{13}$CO counterpart, seems to extend further to -70 \kms in Figure \ref{pv2}b. We note that the $^{12}$CO emission at -40 \kms in Figure \ref{pv2}b is the 3 kpc arm outside the GC and is unrelated with the CO ridges. Detailed velocity distributions of the \twelvecoh and \thirteencol emission are presented in Figures \ref{ap_channel6} and \ref{ap_channel7} in the Appendix.

In order to have a comprehensive view of the complicated features presented above, Figure \ref{schepv}a shows a longitude--velocity diagram toward the CO ridges in \thirteencole. The foreground features have narrow linewidths and are more extended in the sky than those in the GC and are indicated by the green lines at +10 \kmse, 0 \kmse, -30 \kmse and -50 \kms in Figure \ref{schepv}a. Figure \ref{schepv}b shows a schematic diagram of Figure \ref{pv}b. The foreground features in Figure \ref{schepv}a are indicated by the green color. This figure shows that the -35 \kms ridge appears to be connected with the 0 \kms feature at b = 0\fdg5, while the two bridging features are too faint to be seen here. The -50 \kms cloud and the 10 \kms cloud shown by black hatching may be foreground features, but their line-of-sight locations are not certain from the present results alone.

\subsection{Excitation conditions in the 0 \kms feature and the two CO ridges}
The molecular gas in the GC generally shows higher--excitation states reflecting higher temperature than that in the Galactic disk (e.g., Morris \& Serabyn 1996; Torii et al. 2010a; Kudo et al. 2011). The excitation state of the gas can therefore be used as a determinant of whether the CO features are likely in the GC. In the CMZ, \citet{saw1999} found that the average value for the $^{12}$CO $J$=2--1/1--0 intensity ratio is 0.96 $\pm$ 0.01, where their $^{12}$CO data were sampled at 7.5 arcmin intervals with 9--arcmin HPBW for l = $\pm$3\fdg0, b = $\pm$1\fdg0, at a velocity resolution of 2 \kms over a velocity span of $\pm$ 300 \kmse. The ratio is however sensitive to the velocity resolution and spatial sampling rate employed. By using the NANTEN2 OTF data with full sampling on a 1 arcmin grid with 2.6--arcmin HPBW, we find that the $^{12}$CO $J$=2--1/1--0 ratio, R$_{2-1/1-0}$, is $>$ 0.7 in the CMZ, reaching more than 1 in some positions, 0.5--0.6 in the largest velocity features in the ``parallelogramh in the l--v diagram toward the CMZ  (e.g., \citet{bin1991}). Part of the foreground gas, e.g., the CO gas at (l, b, v) = (0.2--1.0, 0.5--0.7, -10 to +10), shows low line ratio like 0.4--0.5, whereas some of the foreground gas toward the $Spitzer$ 8 $\mu$ m features (Figure 22) shows enhanced ratio like 0.8 as seen in the $^{12}$CO feature at (l, b, v) = (-1.7 -- -1.2, -0.8 -- -0.2, -5.1 to +2.5 ). These full results of \rtwoone toward the CMZ region will be published separately (K. Torii et al. 2013b).

\begin{figure*}
\begin{center}
.pdfscale{0.8}
\includegraphics[scale= .70]{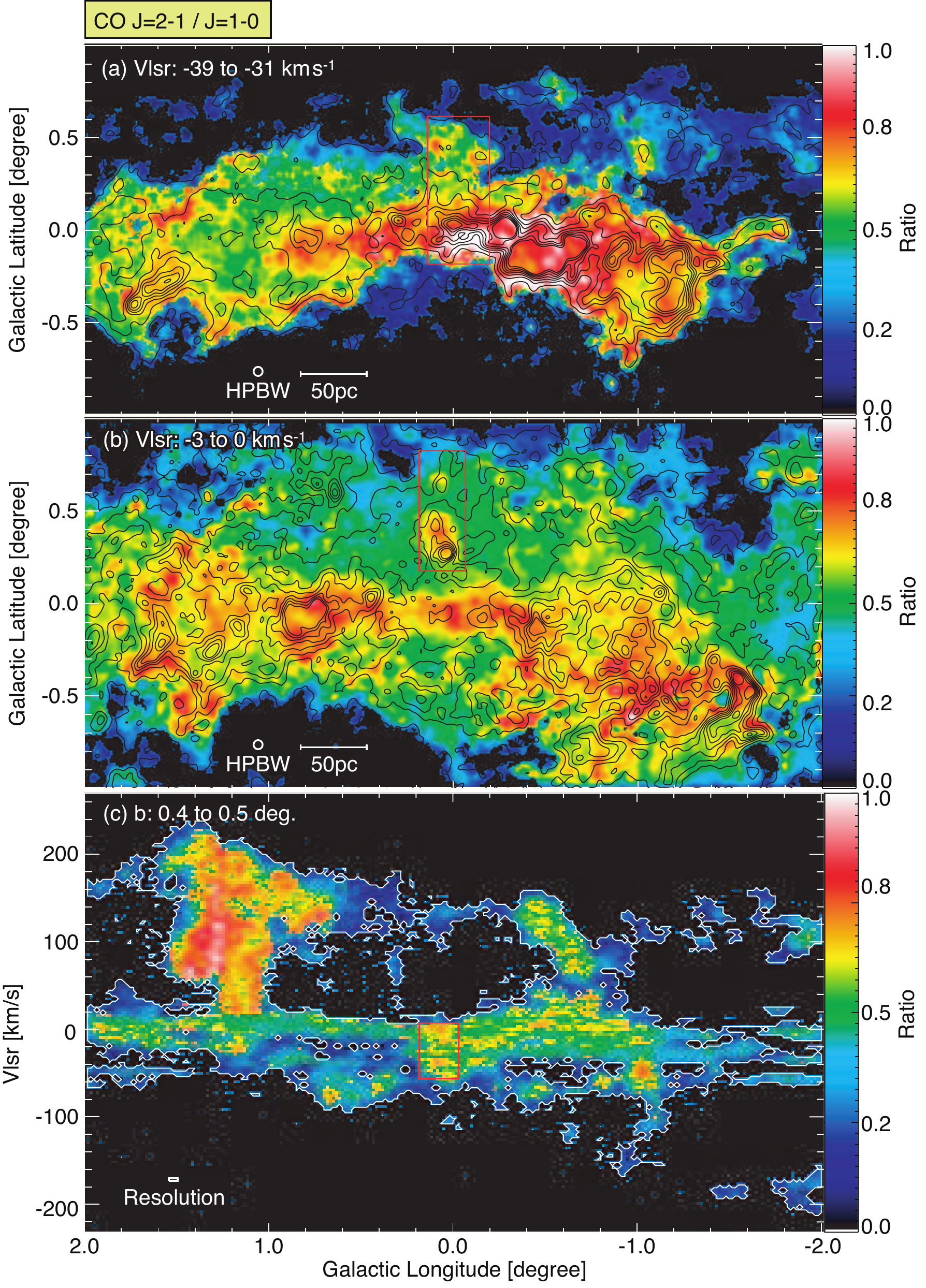}
\caption{(a,b) Intensity ratio distributions between $^{12}$CO $J$=2--1 and $^{12}$CO $J$=1--0 overlaid with $^{12}$CO $J$=2--1 contours. The integration range in $V_{\rm lsr}$ is shown in the figure. Contours are plotted at every 2 K from 2 K. (c) Longitude--velocity diagram of CO ridges. Color shows ratio distributions between $^{12}$CO $J$=2--1 and $^{12}$CO $J$=1--0 overlaid with $^{12}$CO $J$=2--1 contours. The integration range in b is shown in the figure. Red boxes show the region around the two CO ridges, which has a relatively higher line intensity ratio.\label{ratio}}
\end{center}
\end{figure*}

Figure \ref{ratio} shows \rtwoone over the whole region of the present study, including the DHN. Figure \ref{ratio}a shows that the -35 \kms ridge at (l, b) = (-0\fdg 2 to +0\fdg1, 0\fdg4 to 0\fdg5) shows a high line intensity ratio above 0.7 and is conspicuous in the region above b = 0\fdg2. The figure also shows that regions of a high ratio are seen toward the CMZ and the other local $Spitzer$ features having an associated heat source (Figure \ref{sp_fil} in the Appendix). It is possible that the ratio is affected by self--absorption in (l, b, v) = (-0\fdg5 to +0\fdg5, -0\fdg2 to +0\fdg2, -20 \kms to 20 \kmse)  in Figures \ref{ratio}a and \ref{ratio}b. Finally, Figure \ref{ratio}c shows a longitude--velocity diagram of the ratio and that the two CO ridges as well as the bridging features show an enhanced ratio above 0.7, which is different from the other foreground features showing a smaller ratio around 0.3 - 0.5. Accordingly, we suggest that the \rtwoone enhancements in the CO ridges favor their location in the GC.

\subsection{Distance estimation by an analysis of the near--infrared stellar extinction}
In order to make an independent estimate of the distance of the CO features and ridges, we have examined the distribution of near--infrared extinction derived from 2MASS data \citep{dob2011} and have made a comparison with CO distributions.
 
\begin{figure*}
.pdfscale{0.80}
\includegraphics[scale= 0.80]{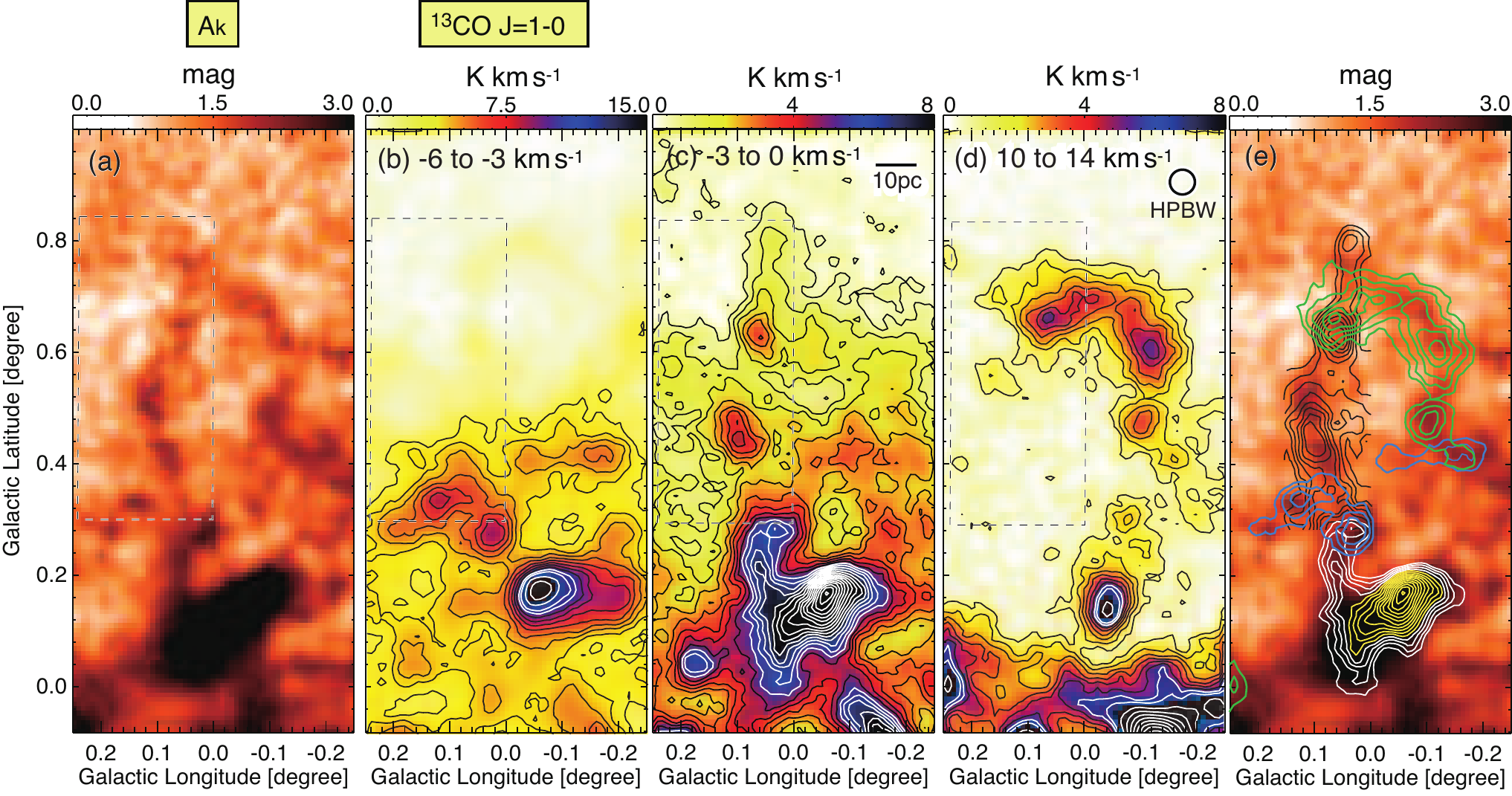}
\caption{(a) A 2MASS $A_K$ extinction map toward the DHN. (b) $^{13}$CO $J$=1--0 distributions integrated from $-6$ to $-3$ km s$^{-1}$. Contours are plotted every 2 K km s$^{-1}$ from 4 K km s$^{-1}$ and then every 4 K km s$^{-1}$ from 34 K km s$^{-1}$. (c) $^{13}$CO $J$=1--0 distributions of the 0 km s$^{-1}$ ridge and feature. Contours are plotted every 0.6 K km s$^{-1}$ from 0.6 K km s$^{-1}$. (d) $^{13}$CO $J$=1--0 distributions of the $10$ km s$^{-1}$ cloud. Contours are plotted every 0.8 K km s$^{-1}$ from 1.0 K km s$^{-1}$. The dashed boxes show the field used for the simulation of the stellar extinction. (e) The same figure as (a) with CO contours in various velocity ranges related to dark filaments in the figure.\label{ext}}
\end{figure*}

Figure \ref{ext}a shows the distribution of A$_{K}$ extinction toward the CO features and ridges. We see significant extinction toward the 0 \kms ridge in addition to several additional features. We identify each feature by comparing it with the CO distribution. Figures \ref{ext}b--\ref{ext}d show the \thirteencol integrated intensity distributions for three velocity ranges of -6 \kms to -3 \kmse, -3 \kms to 0 \kmse, and 10 to 14 \kmse, respectively. We find that the 0 \kms feature (Figure \ref{ext}c) shows a good correspondence with the dark filament at (l, b) = (0\fdg00 to 0\fdg25, 0\fdg3 to 0\fdg85) enclosed by dashed lines in Figure \ref{ext}a. The other filamentary features have counterparts in Figures \ref{ext}b --\ref{ext}d; the extinction feature at  (l, b) = (0\fdg1, 0\fdg3) corresponds to $^{13}$CO emission between -6 \kms to -3 \kmse, those in (l, b) = (-0\fdg15 to 0\fdg1, +0\fdg1 to 0\fdg3) and (-0\fdg1 to +0\fdg1, 0\fdg0 to 0\fdg2) to the $^{13}$CO in -3 \kms to 0 \kmse, and that at (l, b) = (0\fdg25, 0\fdg0) to the $^{13}$CO in 10 \kms to 14 \kmse, as shown by an overlay with the extinction in Figures \ref{ext}e.
 
In order to estimate the distance to the 0 \kms feature, which shows good coincidence with the stellar extinction, we used the stellar population synthesis model, the Besan\'con Galaxy Model, by \citet{rob2012}. This model includes a bar and a thick bulge adopted from the three--dimensional extinction model \citep{mar2006}. The model is based on realistic assumptions about the formation and evolution of four main stellar populations of the Milky Way. It allows one to simulate the stellar content along any given line of sight, and for each simulated star the photometry, kinematics, and metallicity are computed. For each population, a star formation rate history, age, and initial mass function are assumed, which leads to a distribution function in absolute magnitude, effective temperature, and age of the stars. The distribution function is assumed to be controlled by dynamical principles \citep{bie1987} for each population and are tested by means of photometric star counts. We estimated the distances of the components distributed toward the 0 \kms feature (the dashed box in Figures \ref{ext}a--\ref{ext}d) by comparing it with the three--dimensional interstellar extinction distribution by the Besan\'con model. The simulated color--magnitude diagrams are in excellent agreement with the IRSF color--magnitude diagram (Figure \ref{compirsf} in the Appendix), making us confident that we can use this model to predict the distances.
 
\begin{figure*}
\begin{center}
.pdfscale{1.}
\includegraphics[scale= 1.2]{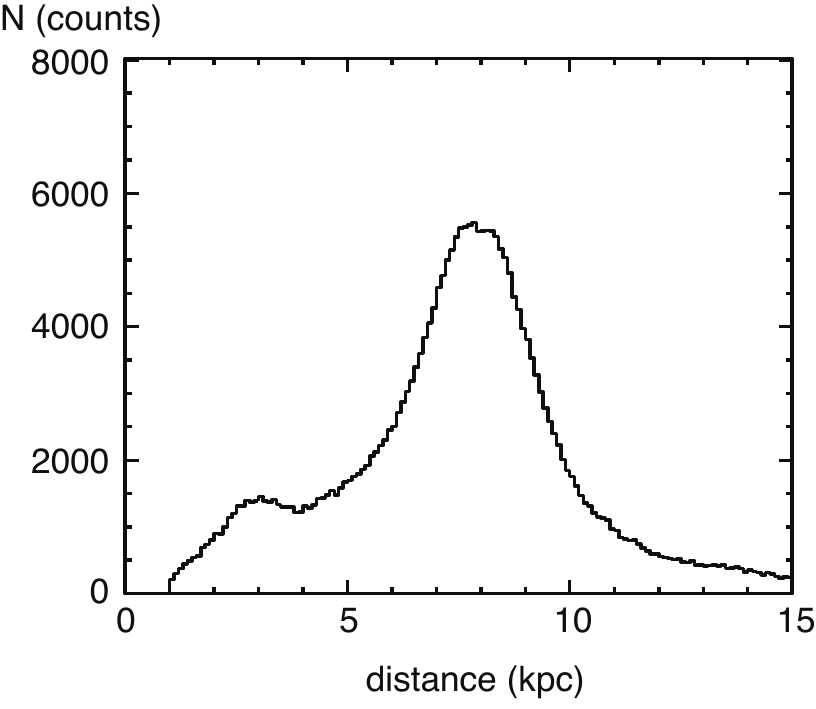}
\caption{Distance plot toward a given field depicted by boxes in Figure \ref{ext}. The vertical axis shows the number of simulated stars. \label{ext2}}
\end{center}
\end{figure*}

The result is shown in Figure \ref{ext2}, which indicates that there are two extinction components toward the direction of the 0 \kms feature and ridge. One is a weak peak located at 2 kpc but, on the other hand, the other has a strong peak at around 8 kpc, the distance to the GC. We also looked at the typical distance distribution of the 3 dimensional--dust extinction map of \citet{mar2006} which shows clearly that in the direction of the 0 \kms feature and ridge most of the extinction is situated at a distance of 8 kpc. \citet{mar2006} also showed the excellent agreement between the dust extinction features and the CO gas, indicating the direct association between the dust and gas features. We therefore suggest that the 0 \kms feature and ridge are located in the GC. However, the latter has an intrinsic dispersion associated with the depth of the bulge itself and we adopt the distance to the 0 \kms feature and ridge as 8 kpc $\pm$2 kpc by taking into account the size of the bulge.

\subsection{Summary of the observational results}
We here summarize the present observational results with NANTEN2 as follows;
\begin{enumerate}
\item We have found two CO features toward the DHN at -35 \kms and 0 \kms which are extended along the DHN. The two CO features are likely associated with the DHN as suggested by the morphological correlation with the infrared nebula. Follow--up observations at three--times higher angular resolutions show detailed correspondence and lend support for this association (K. Torii et al. 2013a submitted). The two features show velocity gradients of 0.4 \kms pc$^{-1}$ in the same sense perpendicular to the long axis of the DHN at a distance of 8 kpc.

\item We have also found two CO ridges at the same two velocities toward the Galactic south of the CO features, that extend from the bottom of the DHN down toward the Galactic plane (Figure \ref{lb13}), although contamination by other features makes these ridges difficult to follow at latitudes below 0\fdg2 in $^{12}$CO (Figure \ref{lb_large}). We note that the CO ridges show spatial correspondence with the $Spitzer$ infrared distribution within l = -0\fdg2 to 0\fdg2 and b above 0\fdg3, where contamination is not so heavy. In particular, correspondence with CO emission is good for the infrared DHN and its filament directed toward the Galactic plane (Figure \ref{lb13}). It is apparent that the 0 \kms ridge is connected with the 0 \kms feature, as shown in the latitude--velocity diagram (Figure \ref{pv}). On the other hand, connection of the -35 \kms feature with the -35 \kms ridge is not as clear in the latitude--velocity diagram at b = 0\fdg4 -- 0\fdg6 (Figure \ref{pv}) as it is for the 0 \kms ridge. Nevertheless, the -35 \kms ridge has a large velocity span of 20 \kms in $^{12}$CO (Figures \ref{pv}a and \ref{pv2}b), suggesting that the ridge is located in the GC. The two CO ridges appear to be connected in space and velocity by kinematically broad CO features at b = 0.5 degree (gthe connecting featureh in Figures \ref{pv}, \ref{cnct}b, \ref{pv2}b and \ref{schepv}b) and two broad features within l = -0\fdg1 -- 0\fdg1 and b = 0\fdg1 to 0\fdg3 (Figures \ref{pv2}b and \ref{pv2}c), suggesting that the two ridges are physically linked with each other. 

\item Parts of the two CO ridges show a relatively high line intensity ratio \twelvecoh /$J$=1--0 of 0.6 -- 0.7, toward (l, b) = (0\fdg1, 0\fdg5) at -35 \kms (Figure \ref{ratio}a), toward (l, b) = (0\fdg1 to 0\fdg2 , 0\fdg3 to 0\fdg5) at 0 \kms (Figure \ref{ratio}b), and (l, b) = (-0\fdg1 to +0\fdg1, 0\fdg4 to 0\fdg5) at between -50 \kms and +10 \kms (Figure \ref{ratio}c). The ratio is higher than that of the foreground gas, 0.3 -- 0.5, and is likely due to the higher temperature of the molecular gas; this is consistent with our hypothesis that the ridges are located in the GC, where the molecular gas as a whole shows higher temperatures than in the Galactic disk (e.g., Morris $\&$ Serabyn 1996). 

\item We have examined extinction in the K band toward the 0 \kms feature and the 0 \kms ridge and have shown that the extinction is well modeled by the Besan\'con Galaxy Model, for a distance of 8 kpc $\pm$ 2 kpc to the 0 \kms ridge, which appears clearly in the star count image. This value is consistent with a location in the GC.

\end{enumerate}

In summary, we suggest that all the ridges and features at -35 \kms and 0 \kms are probably located in the GC and it is possible that these ridges and features are physically linked altogether. More robust verification of their physical connection must await observations of higher--$J$ CO transitions that are less contaminated by lower--excitation foreground gas.

\section{Discussion}
Since the discovery of the DHN the physical and kinematical properties such as mass and velocity of the DHN have been unknown, hampering our understanding of its origin, based so far only on morphology. One suggestion is that the CND (circum--nuclear disk) which is rotating around the super--massive black hole, Sgr A*, is driving the DHN \citep{mor2006}. The DHN was suggested to be of magnetic origin by these authors, since its double helix distribution is similar to what is expected to be formed via torsional Alfve\'n waves, which had been discussed to interpret helical CO distributions in the L1641 cloud and  the $\rho$ Oph streamers (e.g., Uchida et al. 1990, 1991). On the other hand, \citet{law2010} and \citet{tsu2010} suggested that the DHN is located at the top of the polarized radio plume, which is driven by the Radio Arc. The origin of the Radio Arc is not yet understood and the energy that can be released from the Radio Arc is unclear; the total magnetic energy stored in the Radio Arc may be on the order of 10$^{52}$ erg for an average field of mG and volume of 60 pc $\times$ 15 pc $\times$ 15 pc. In this scenario, the origin of the DHN shape remains unexplained. 

The present CO results have opened an opportunity to explore the dynamical properties of the DHN for the first time. In the following, we compared the relevant data with the present CO images in order to elucidate the mechanisms underlying the DHN.

\begin{figure*}
\begin{center}
.pdfscale{1.}
\includegraphics[scale= .80]{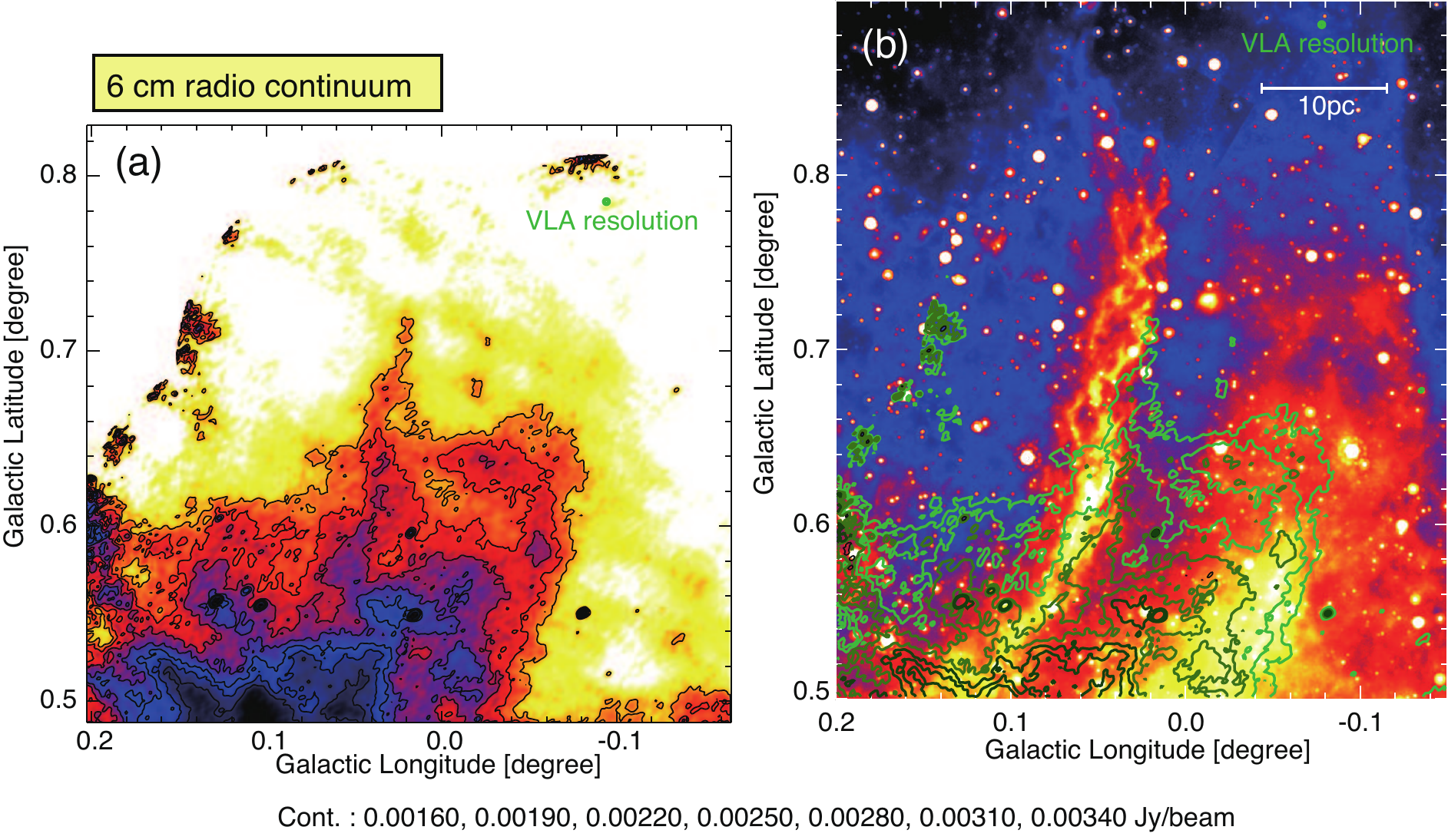}
\caption{(a) 6cm radio continuum distribution given by \citet{law2008}. (b) Contours of (a) superposed on the $Spitzer$ 24 $\mu$m image.\label{law}}
\end{center}
\end{figure*}

Figure \ref{law}a shows the 6cm radio continuum emission at $\sim$10 arcsec resolution. Figure \ref{law}b shows 6cm radio continuum contours overlaid on the $Spitzer$ 24 $\mu$m distribution \citep{law2008}. It is seen that the radio continuum emission b $\ge$ 0\fdg6 traces part of the infrared distribution including the DHN and its connected filament as well as another broad filament 0\fdg1 to the west of the DHN. The radio emission toward the DHN is likely non--thermal since it is polarized \citep{tsu2010}. Similar radio continuum observations toward the whole extent of the CO ridges down to b $\sim$ 0\fdg0 are required to better understand the relationship of the radio continuum with the present CO distributions. 

Figures \ref{lb13}c--\ref{lb13}d show the distributions of the $Spitzer$ 24 micron emission, the radio polarized plume and the CO ridges in $^{13}$CO $J$=1--0. The locations of the Radio Arc and the CND are also compared with the $Spitzer$ image in Figures \ref{lb13}c--\ref{lb13}d. The 0 \kms ridge is along the polarized plume but becomes bent toward the CND at b = 0\fdg5. The eastern leg of the -35 \kms ridge seems to be elongated toward the Radio Arc but the radio polarized plume is located in the east beyond the -35 \kms ridge. The western leg of the -35 \kms ridge is extended toward the CND. The leg of the -35 \kms ridge toward (l, b) = (0\fdg05 , 0\fdg00) is contaminated by the strong CO emission with large velocity width (see e.g., Figure \ref{lb13}a). To summarize, we find that the present ridges show signs favoring association with the CND rather than with the radio polarized plume, whereas a better confirmation of the association is desirable by higher--$J$ CO transitions with much less contamination than lower--$J$ CO transitions.

By sensitive CO observations with NANTEN2 we have discovered at least seven filamentary CO features which extend from the CMZ nearly vertically to the Galactic plane (a fuller account will be presented by Enokiya et al. (2013)). The filaments are associated with more diffuse CO halos having 10$^5$ \msun surrounding the CMZ. Figure \ref{mfs} gives a schematic of the filamentary CO features. The two CO ridges below the DHN are two of these CO filaments. We have shown that the observational signatures of the present CO ridges are consistent with being located in the GC (Section 3). The filamentary CO filaments have velocities from -200 \kms to 160 \kms and large linewidths of more than 60 \kms except for the present CO features. It is likely from a kinematic point of view that all of these features are located in the GC because of their large velocities and velocity spans. Two molecular filaments above Sgr C discussed in \citet{uch1994} seem to be associated with infrared sources, AFGL5376 and other infrared filementary structures observed with $Spitzer$ \citep{sto2006}. The total mass and kinetic energy involved in all the filaments amounts to 10$^6$ \msun and 10$^{52}$ erg, respectively. \citet{uch1994} discussed that these filaments and the DHN are similar in shape and are of shock--related origin. Their kinematics in CO emission is however quite different. In particular, the peak velocities of the DHN features are -35 \kms and 0 \kmse, much lower than that of the two filaments toward AFGL5376, whose velocity ranges from 40 to 160 \kmse, suggesting that the DHN is of different origin from the rest. The possibility of a relation between the DHN and the other infrared filaments will be discussed separately (Enokiya et al. 2013b).

The CO filaments at high z around 100 pc require some energy source and mechanism for their formation against the strong stellar gravity in the inner 100 pc of the GC. Active high--mass star formation may provide such energy via stellar winds, supernova explosions etc. (e.g., Crocker et al. 2013). Except for the CO features and ridges having low velocities close to zero, the other filamentary features have velocities similar to that of the ``parallelogramh in the l--v diagram at around 100 -- 200 \kmse, outside the velocity range of the Sgr A, Sgr B2 and Sgr C giant molecular clouds, whose velocity range from -50 \kms to 100 \kms (see e.g., Figure \ref{ap_channel3}). These three giant molecular clouds are known to be active in star formation, while the ``parallelogramh in the l--v diagram shows no sign of active star formation (see Section 3.4). High--mass stars may possibly accelerate the gas (e.g., \citet{cro2012}), but the highest velocity observed, more than 100 \kms, is too large to be explained by such a process.
Another possible mechanism to lift up the molecular gas is the Parker instability \citep{par1966}. The three molecular loops at kpc separation from the GC, loops 1, 2, and 3, discovered by NANTEN CO observations \citep{fuk2006, tor2010a, tor2010b, fuj2009} show that the Parker instability is operating to levitate molecular gas vertically to the Galactic plane under the magnetic field of 100 $\mu$G or higher. It is probable that the magnetic field strength is even higher in the inner 100 pc than at 1 kpc of the GC, where loops 1, 2 and 3 are located. The typical kinetic energy involved in these loops is 10$^{51}$ erg \citep{fuk2006} and the Parker instability offers an alternative mechanism to create the filaments except for the DHN. According to these previous works, the features created by the Parker instability show very large linewidth like 50--100 \kms, which is much larger than the velocity dispersion of the DHN. We therefore do not consider that the DHN is created by the Parker instability, whereas the other filamentary features having large linewidths are well explained by the Parker instability.

According to the above discussion, the CND seems to be a plausible candidate for the driving engine in creating the DHN via torsional Alfv\'enic waves as originally suggested \citep{mor2006}. The magnetic gtowerh above the central massive black hole in Galactic centers is a common signature of MHD numerical simulations of a magnetized rotating nuclear disk of a galaxy and they originate from the twisted magnetic field lines around the rotating disk surrounding the central super--massive black hole \citep[e.g., ][]{mac2009}. A possible picture along this line is that magnetic fields anchored to the two diametrically--opposite parts of the disk are twisted to form a double helix pattern of the magnetic fields, which creates the infrared DHN and ridges. Such a model requires realistic three--dimensional MHD numerical simulations for the appropriate parameters of the CND and should be compared with the observations in detail in the future. This is however beyond the scope of the present observational paper, and we simply describe a basic scenario for such a possible model below.

The model may be applied to the CO feature(s) and the CO ridge(s) as the molecular gas associated with the magnetic tower, and we show parameters for the both cases. Since the magnetic tower is hypothesized to be rising from the CND in response to magnetic pressure \citep{mac2009}, the ambient molecular and/or atomic gas will be driven and possibly dragged by the rising magnetic field lines. We find that neutral gas in the halo may serve as the mass reservoir in the interaction. For a typical rotation speed of 80 \kms at 2 pc radius of the CND \citep{oka2011}, the rotation period of the field lines is 0.15 Myr one rotation is supposed to form one turn of the double helix, whose length corresponds to 15 pc. This gives a total number of turns around 8 for the total length of the DHN and ridges, $\sim$120 pc, about 20 \% of which corresponds to the DHN. A rising velocity of the filament is estimated to be 100 \kms through dividing 15 pc by 0.15 Myr. The timescale of the DHN formation only is estimated to be $\sim$0.2 Myr for a length observed as the DHN, $\sim$ 20 pc, or 0.8 Myr - 1.0 Myr ago from the current epoch, and $\sim$1.2 Myr if the whole system of 120 pc length is considered. We note that the observed velocity gradients in the -35 \kms and 0 \kms features are qualitatively consistent with that of the CND \citep{oka2011}. This is interesting, even if the velocity gradient is two orders of magnitude smaller than that on the CND, because the angular momentum of the magnetic tower will be significantly lost by the interaction with the ambient gas.

Total kinetic energy of the system is estimated to be 10$^{47}$ erg (only for the 0 \kms DHN) to 10$^{51}$ erg (including the two velocity CO features and ridges) by considering only the lateral expanding motion perpendicular to the rotation axis of the CND. The mass accretion rate required to release such energy is roughly estimated to be $\sim$ 10$^{-5}$ \msunpyr at 2 pc radius in the CND for the DHN only, and $\sim$ 10$^{-2}$ \msunpyr for the whole features and ridges if a complete conversion of the released energy into the kinetic energy is assumed, where we take the mass of the Sgr A* to be 4$\times$10$^6$ \msun \citep{ghe2008, gil2009}. This mass accretion rate is not uncomfortably large. It is to be noted that this mass accretion rate does not directly apply to the accretion to Sgr A* itself, because most of the accreting mass in the CND will probably be ejected outward at much large radii as the magnetic tower.

We finally note that similar jet--like elongated molecular clouds are observed toward Westerlund 2 (Fukui et al. 2009, Furukawa et al. 2013 submitted), and SS433 and MJG348.5 \citep{yam2008}. The possible driving source includes a compact stellar remnant like a black hole or a neutron star which may be related to the gamma--ray emission. Based on these observations it is suggested that the interaction between high--energy jet with the surrounding neutral clouds, either in molecular and/or dense atomic gas, is able to form straight and long molecular gas along the jet axis \citep[][]{yam2008, fuk2009}. We admit that the magnetic tower proposed here is not the high energy jet in SS433. Nonetheless, the basic physics in the interaction shares similarities between the two in particular in the dynamical interaction with the ambient gas. Asahina et al. (2013) successfully developed MHD numerical simulations aiming at explaining the physical processes involved in these jet--like clouds. Such a model may offer a possible explanation to form the present CO features/ridges.

\section{Conclusions}
We have discovered two candidate molecular features associated with the infrared Double Helix Nebula (DHN) in the distribution of the \twelvecohe, \twelveco and \thirteenco $J$=1--0 emission obtained with NANTEN2 in the GC. Two CO features at velocities of -35 \kms and 0 \kms correspond well to the infrared distribution of the DHN. This association is reinforced by follow--up observations at three--times higher angular resolutions (K. Torii et al. 2013a submitted). We have also found two elongated CO ridges at these two velocities oriented vertically to the Galactic plane over 0.8 degree. The two ridges are linked by broad velocity features and are likely connected physically with each other. The physical connection between the CO DHN and the two CO ridges is possible because of the coincidence in space and velocity, while contamination by other CO features in the same direction hampers to firm establishment of an association between the two CO features and ridges. It is likely that the CO counterparts of the DHN are in the GC. An analysis of the K band extinction shows a good coincidence with the 0 \kms ridge and a modeling of the three--dimensional stellar distribution yields a distance of 8$\pm$2 kpc to the 0 \kms ridge. The large velocity span of the -35 \kms ridge, $\ge$ 20 \kmse, favors its location in the GC. We therefore suggest that the two ridges are located in the GC. We suggest that further observations in higher $J$ transitions of CO etc. may provide a more convincing evidence for their locations. The molecular mass of the DHN is estimated to be $\sim$ 10$^4$ \msune, while that of the ridges is $\sim$10$^5$ \msune. The kinetic energies involved in the DHN and the ridges are roughly estimated to be 10$^{47}$ erg and 10$^{51}$ erg, respectively for assumed minimum velocity spans of 1 \kms to 20 \kmse. The placement and orientation of the ridges in $^{13}$CO suggests that the 0 \kms ridge may be linked to the CND. We discuss a possibility that the DHN has its origin at the CND and was created by a magnetic phenomenon incorporating torsional Alfv\'en waves launched from the CND as proposed by \citet{mor2006}. If we assume the CND as the origin of driving the DHN, a mass accretion rate in the order of 10$^{-2}$ \msunpyr -- 10$^{-5}$ \msunpyr at 2 pc radius of the CND will explain the kinetic energy of the DHN as due to a past ejection event $\sim$1 Myr ago, or alternatively, the DHN and the ridges as a whole may be driven by a continuous ejection process over the last $\sim$ 1 Myr.\\

NANTEN2 is an international collaboration of 10 universities: Nagoya University, Osaka Prefecture University, University of Cologne, University of Bonn, Seoul National University, University of Chile, University of New SouthWales, Macquarie University, University of Sydney, and University of ETH Zurich. This work is financially supported by Grant-in-Aid for Scientific Research (KAKENHI) from Japan Society for the Promotion of Science (JSPS; Nos. 24224005, 23403001, 23006148-01, 22740119, 22540250, 22244014, and 23740149-1), young researcher overseas visits program for vitalizing brain circulation (no. R2211) from JSPS, and the Grant-in-Aid for Nagoya University Global COE Program, "Quest for Fundamental Principles in the Universe: From Particles to the Solar System and the Cosmos", from the Ministry of Education, Culture, Sports, Science and Technology of Japan (MEXT). This work is based in part on archival data obtained with the $Spitzer Space Telescope$, California Institute of Technology under a contract with NASA. Support for this work was provided by an award issued by JPL/Caltech. This publication makes use of data products from the Two Micron All Sky Survey, which is a joint project of the University of Massachusetts and the Infrared Processing and Analysis Center/California Institute of Technology, funded by the National Aeronautics and Space Administration and the National Science Foundation. 

\acknowledgments

\clearpage


\clearpage

\clearpage

\clearpage

\clearpage

\clearpage

\clearpage

\clearpage

\clearpage

\clearpage

\clearpage

\clearpage

\clearpage

\clearpage

\begin{figure*}
.pdfscale{1.05}
\includegraphics[scale= 1.05]{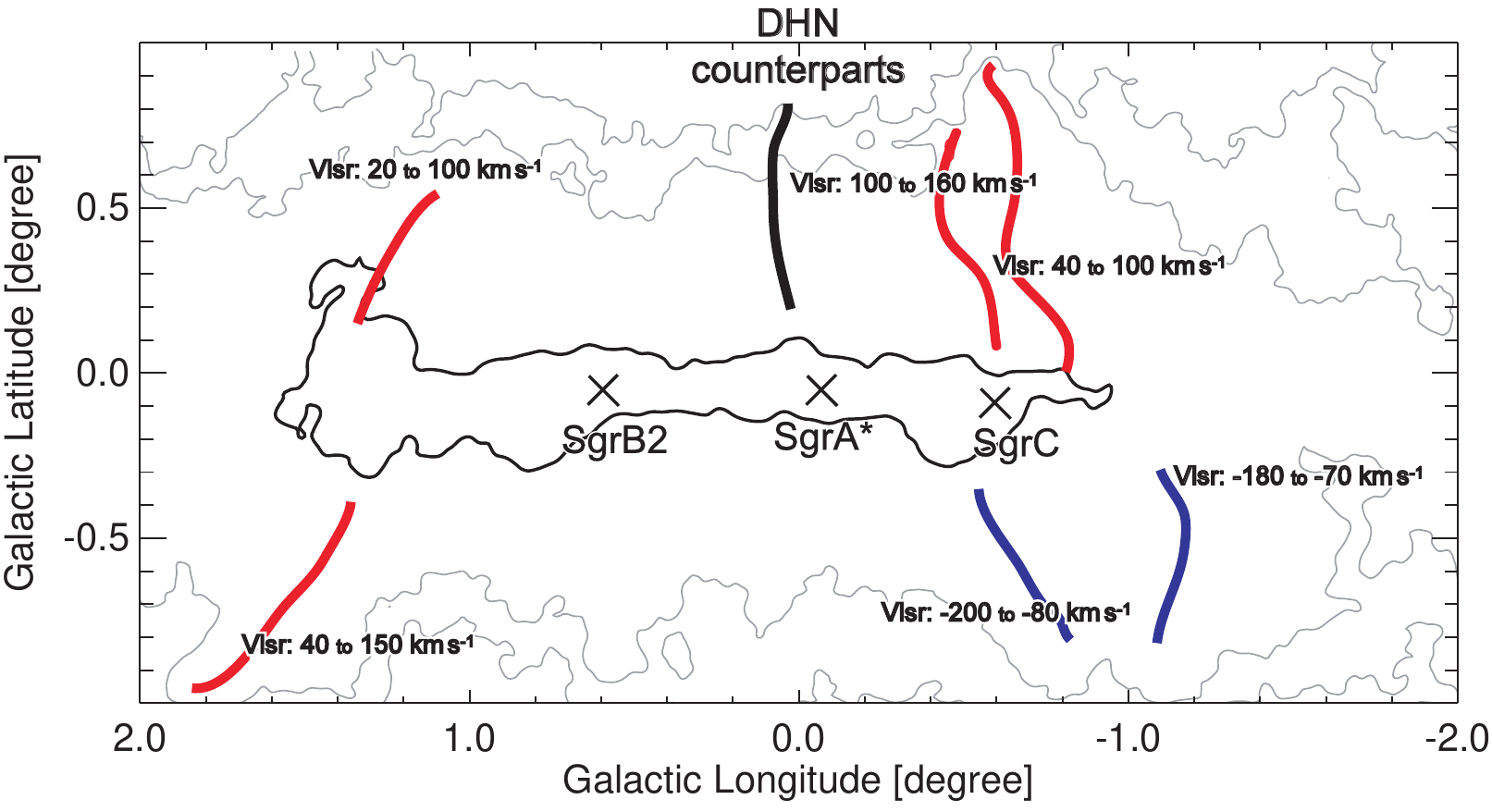}
\caption{Schematic image of filament--like molecular features. Black contour shows the main molecular complex of the CMZ. Gray contours show diffuse high--latitude gases first detected by our observation. Each molecular filament feature can be seen in the detailed large--scale velocity channel maps of Figure \ref{ap_channel1}-\ref{ap_channel3} of the Appendix.\label{mfs}}
\end{figure*}
\clearpage

\begin{figure*}
.pdfscale{.90}
\includegraphics[scale= .90]{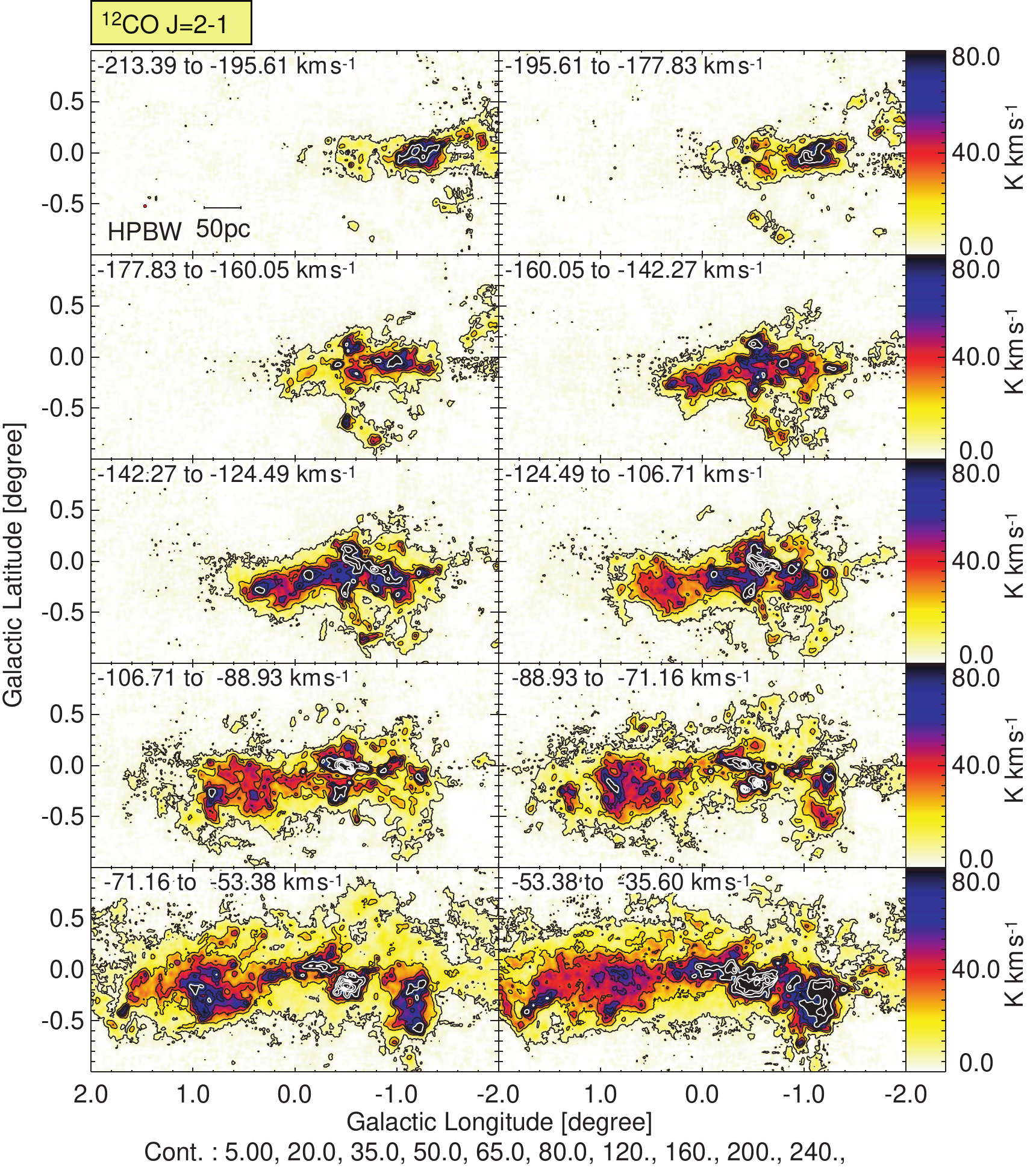}
\caption{Velocity channel distributions of the $^{12}$CO $J$=2--1 emission integrated every 16.51 \kms between -213.4 and -35.6 \kmse. The contour values are indicated at the bottom of the figure.\label{ap_channel1}}
\end{figure*}

\begin{figure*}
.pdfscale{.90}
\includegraphics[scale= .90]{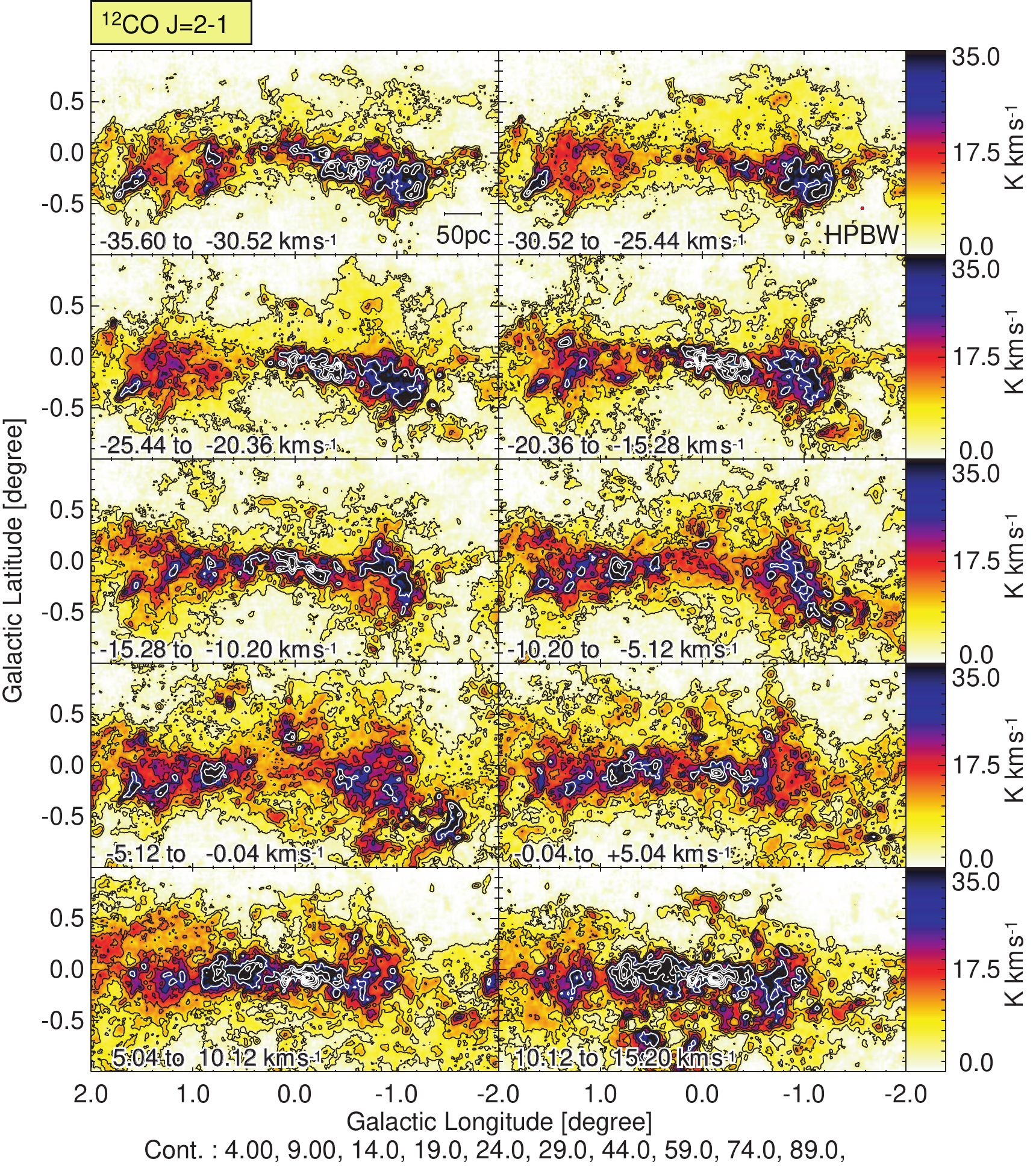}
\caption{Velocity channel distributions of the $^{12}$CO $J$=2--1 emission integrated every 5.08 \kms between -35.6 and $+15.2$ \kmse. The contour values are indicated at the bottom of the figure.\label{ap_channel2}}
\end{figure*}

\begin{figure*}
.pdfscale{.90}
\includegraphics[scale= .90]{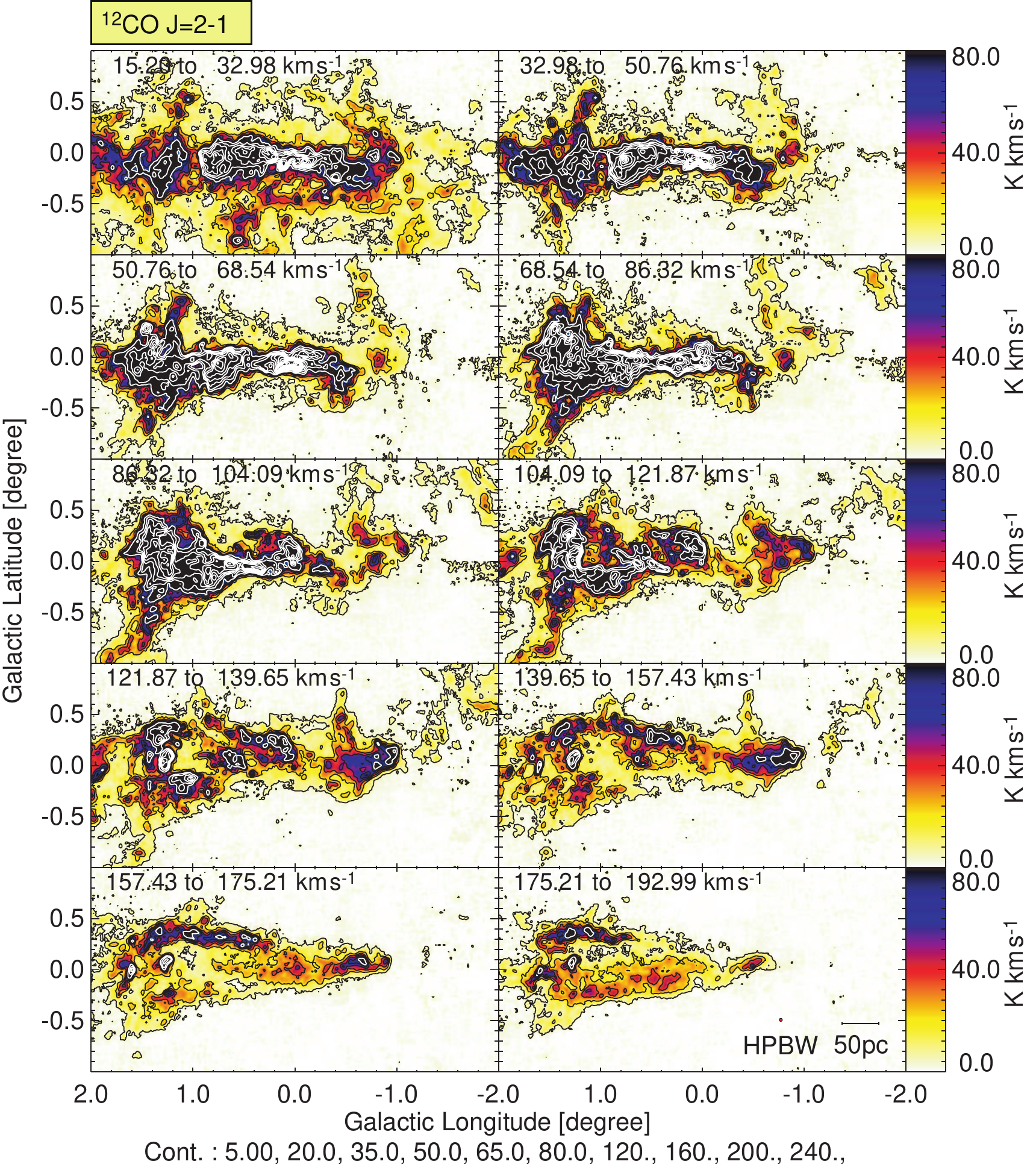}
\caption{Velocity channel distributions of the $^{12}$CO $J$=2--1 emission integrated every 16.51 \kms between 15.2 and 193.0 \kmse. The contour values are indicated at the bottom of the figure.\label{ap_channel3}}
\end{figure*}

\begin{figure*}
\begin{center}
.pdfscale{.80}
\includegraphics[scale= .80]{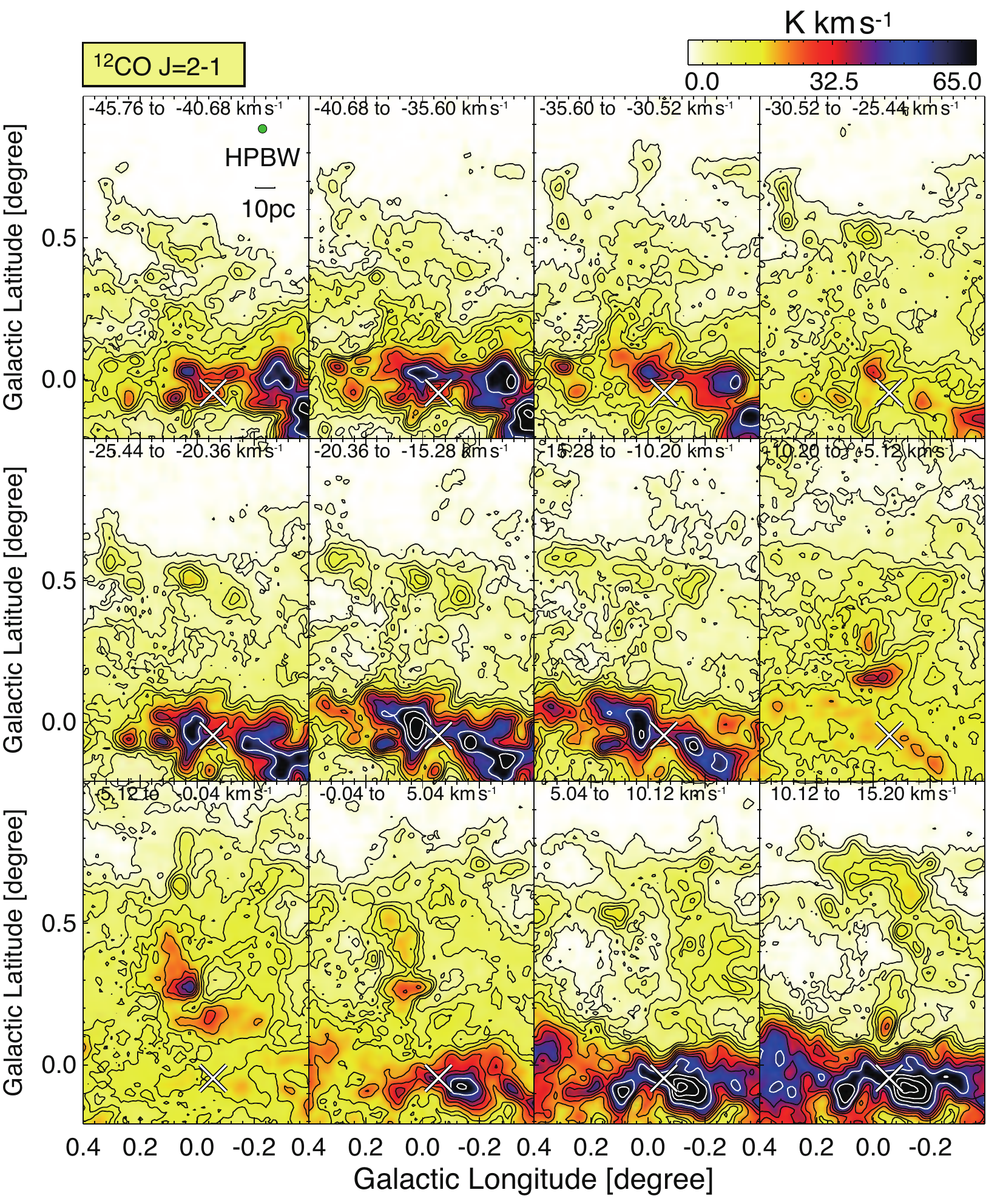}
\caption{Velocity--channel distribution of the $^{12}$CO $J$=2--1 emission integrated every 5.08 km s$^{-1}$ from
-45.76 to +15.20 km s$^{-1}$. Contours are plotted at every 3.2 K km s$^{-1}$ from 2.2 to 11.8 K km s$^{-1}$,
every 6.5 K km s$^{-1}$ from 15.0 to 47.5 K km s$^{-1}$ and every 18.0 K km s$^{-1}$ from 54.0 to 126.0 K km s$^{-1}$. White crosses indicate the position of SgrA*.\label{ap_channel4}}
\end{center}
\end{figure*}

\begin{figure*}
\begin{center}
.pdfscale{.80}
\includegraphics[scale= .80]{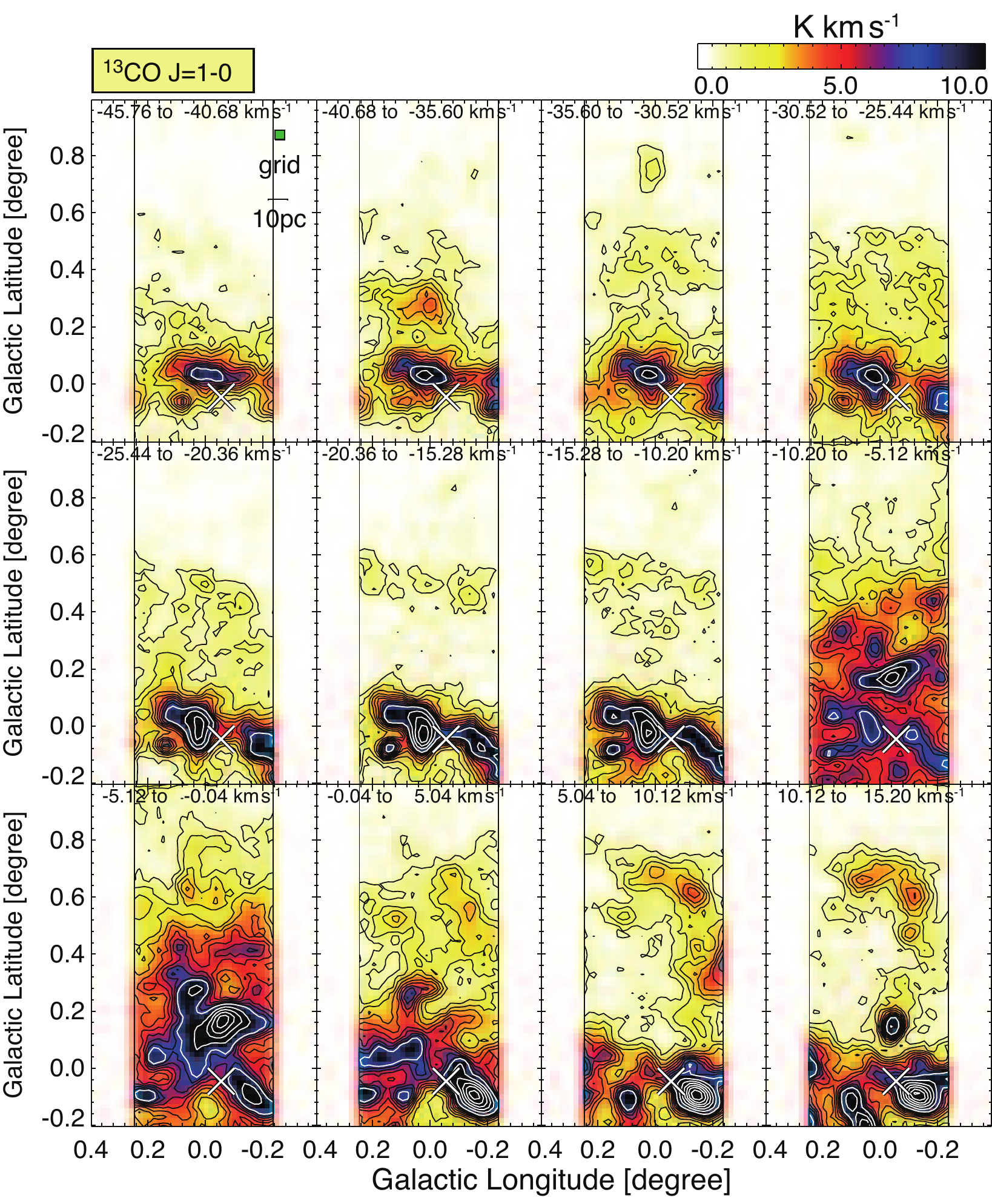}
\caption{Velocity--channel distribution of the $^{13}$CO $J$=1--0 emission integrated every 5.08 km s$^{-1}$ from -45.76 to +15.20 km s$^{-1}$. Contours are plotted at every 0.8 from 0.8 K km s$^{-1}$ to 7.2 K km s$^{-1}$ and every 3.0 K km s$^{-1}$ from 8.0 K km s$^{-1}$. White crosses indicate the position of SgrA*.\label{ap_channel5}}
\end{center}
\end{figure*}

\begin{figure*}
.pdfscale{.90}
\includegraphics[scale= .90]{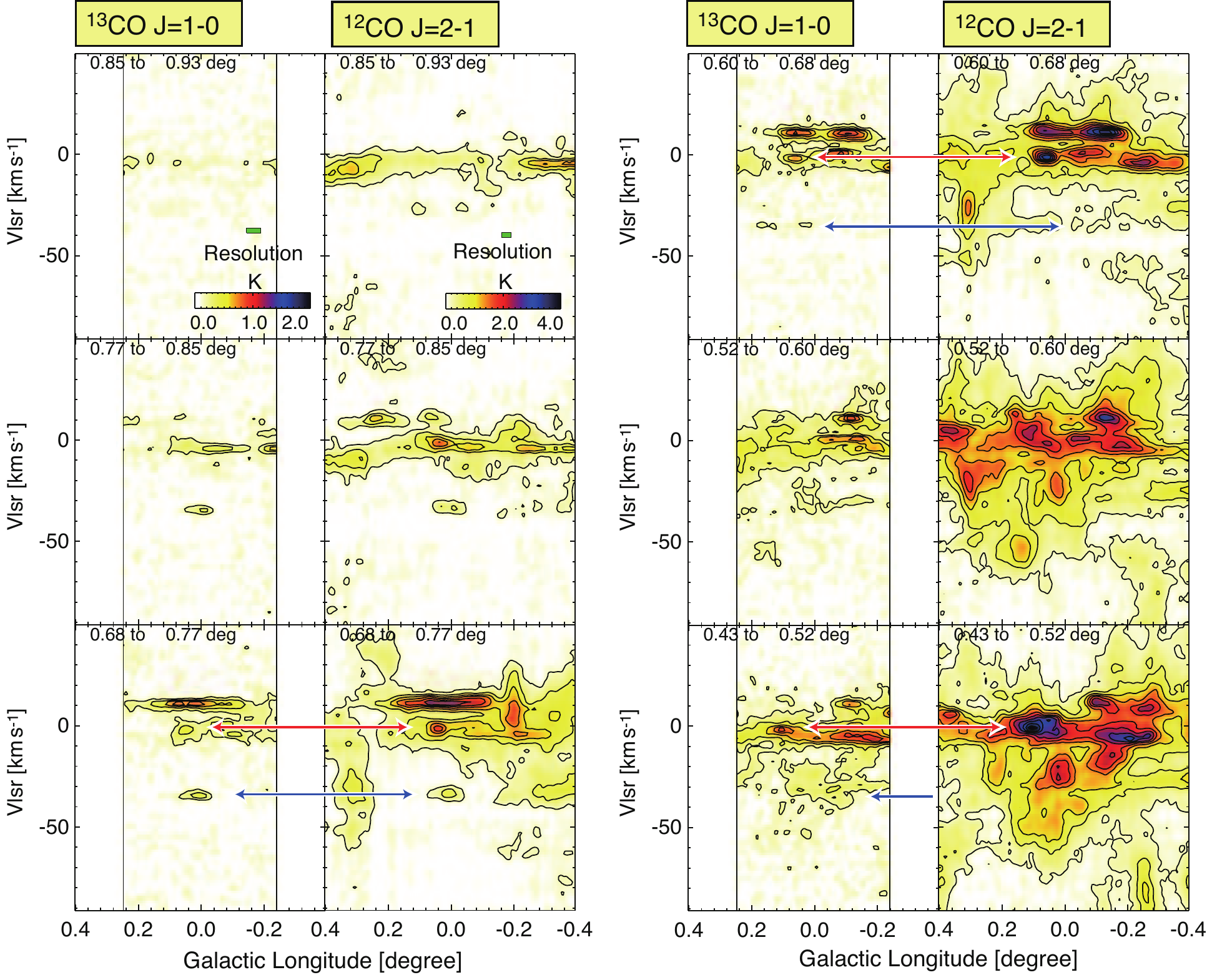}
\caption{Latitude--velocity channel distribution of the $^{13}$CO $J$=1--0 (at the left) and $^{12}$CO $J$=2--1 emission (at the right) integrated in 0\fdg08 latitude intervals from 0\fdg43 to 0\fdg93. Red and Blue arrows show the 0 and -35 \kms feature/ridge, respectively.\label{ap_channel6}}
\end{figure*}

\begin{figure*}
.pdfscale{.90}
\includegraphics[scale= .90]{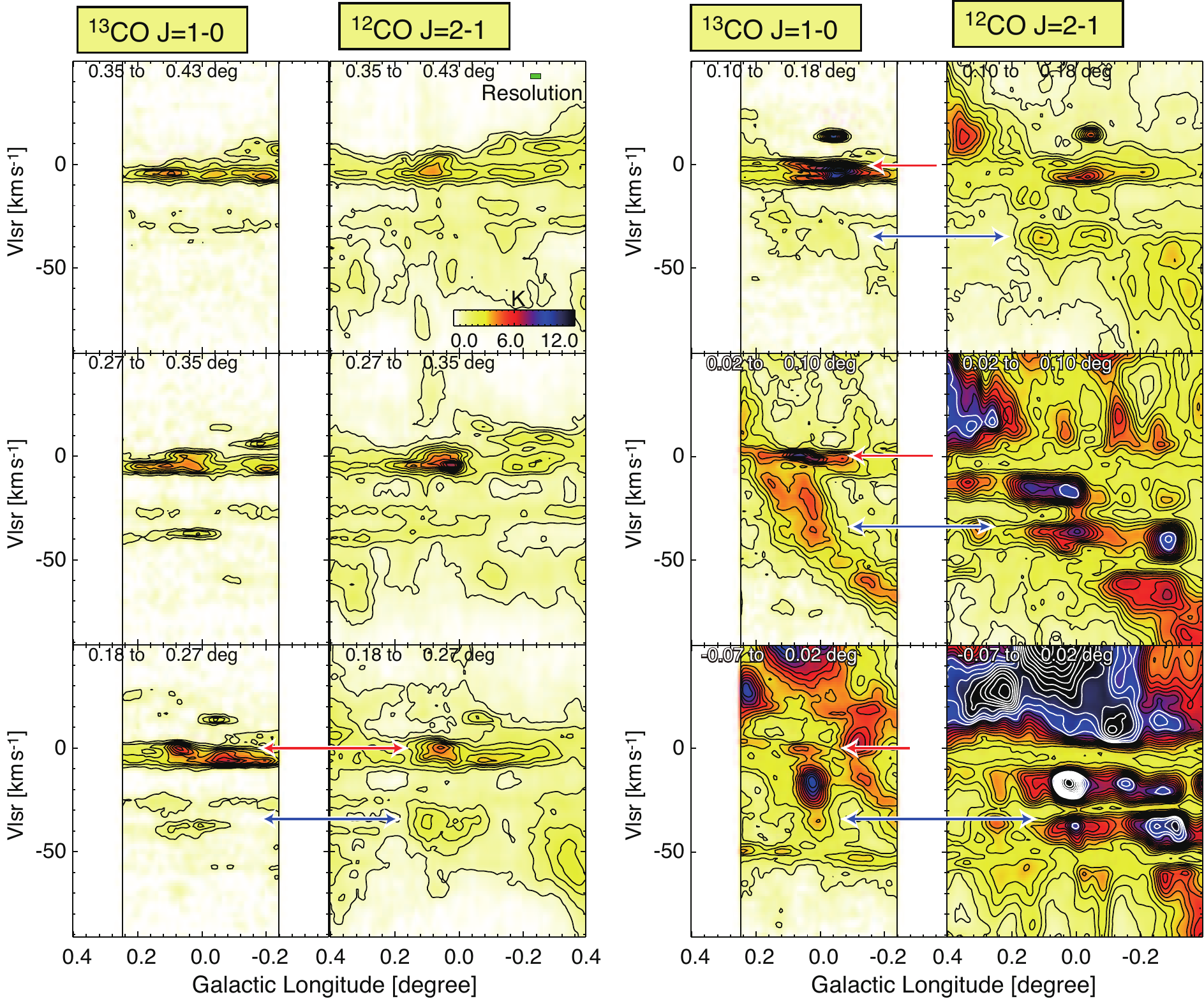}
\caption{Latitude--velocity channel distribution of the $^{13}$CO $J$=1--0 (at the left) and $^{12}$CO $J$=2--1 emission (at the right) integrated in 0\fdg08 latitude intervals from -0\fdg07 to 0\fdg43. Red and Blue arrows show the 0 and -35 \kms feature/ridge, respectively.\label{ap_channel7}}
\end{figure*}

\begin{figure*}
\begin{center}
.pdfscale{.80}
\includegraphics[scale= .80]{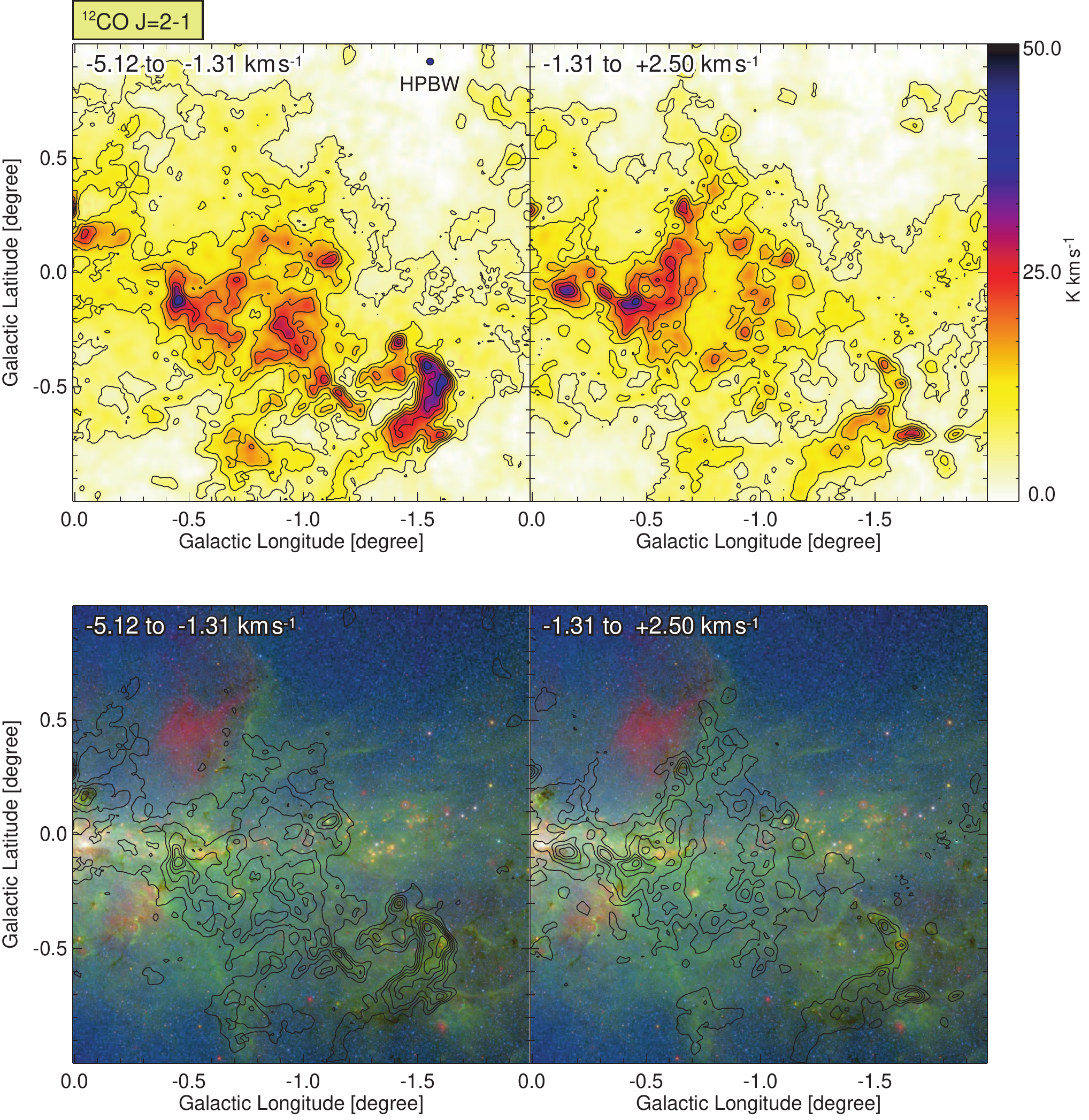}
\caption{(a)-(b) Velocity channel distributions of $^{12}$CO $J$=2--1 emission. The integration range in $V_{\rm lsr}$ is shown in the figures. Contours are plotted at every 5 K~\kms from 5 K~\kms . (c)-(d) Superpositions of CO contours shown in (a) and (b) on the $Spitzer$ composite images including 3.6 $\mu$m (blue), 8$\mu$m (green) and 24$\mu$m (red).\label{sp_fil}}
\end{center}
\end{figure*}

\begin{figure*}
\begin{center}
.pdfscale{.80}
\includegraphics[scale= .80]{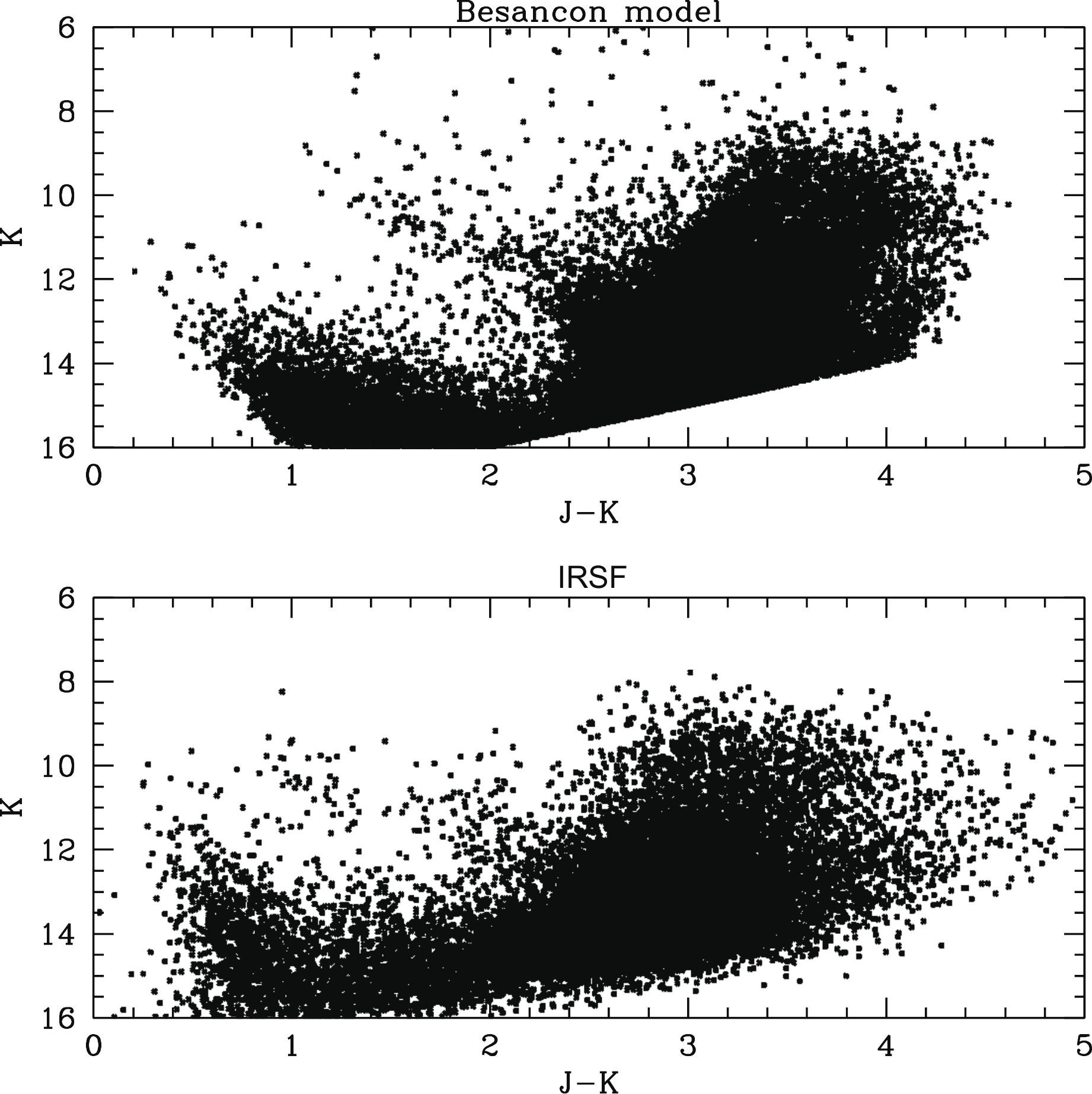}
\caption{Comparison between the observed $J-K$ vs. $K$ diagram (lower panel) and the synthetic color-magnitude diagram from the Besan\c{c}on stellar population synthesis is model \citep{rob2012}. The foreground branch at 2\,kpc and the galactic Bulge red giant branch at 8\,kpc are clearly visible. \label{compirsf}}
\end{center}
\end{figure*}

\clearpage
\appendix
Figures \ref{ap_channel1}--\ref{ap_channel3} show the velocity--channel distributions of the \twelvecohe. Several molecular vertical filaments and also the 0 \kms and -35 \kms ridge can be seen in these maps. Summary of these molecular filaments is shown in Figure \ref{mfs}. We will publish details of molecular filaments in another paper (R. Enokiya et al. 2013b in preparation).
Figures \ref{ap_channel4} and \ref{ap_channel5} show the velocity--channel distributions of the $^{12}$CO $J$=2--1 and $^{13}$CO $J$=1--0 emission. They show the detailed distributions of the discovered two CO features and ridges and the connecting feature. While the $^{12}$CO $J$=2--1 emission in $|b| <$ 0\fdg2 is strongly affected by absorption especially at the velocity range around 0 km s$^{-1}$, the $^{13}$CO $J$=1--0 emission clearly describes the gas distributions not affected by absorption.
Figures \ref{ap_channel6} and \ref{ap_channel7} show latitude--channel distributions of the $^{12}$CO $J$=2--1 and $^{13}$CO$J$=1--0 emission. Diffuse and broad velocity features seen in $^{12}$CO $J$=2--1 at (l, b, v)$\sim$ (-0\fdg2 -- 0\fdg2, -0\fdg07 -- 0\fdg60, -50 \kms -- 0 \kms) fills the interspace between two molecular ridges at v of $-35$ and 0 km s$^{-1}$. Emission from foreground clouds can be seen as narrow velocity features in these figures.
Figure \ref{sp_fil} shows velocity--channel distributions of foreground molecular cloud with high intensity ratio which corresponds to $Spitzer$ filament. 
Figure \ref{compirsf} is comparison between color--magnitude diagram of the Besan\'con model and observational data of IRSF.

\clearpage

\clearpage

\clearpage

\clearpage

\clearpage

\clearpage

\clearpage

\clearpage

\end{document}